\documentclass[12pt,preprint]{aastex}

\begin{document}

\title{Peculiar Velocity Limits from Measurements of the Spectrum of the Sunyaev-Zel'dovich Effect in
Six Clusters of Galaxies}

\author{B.A. Benson\altaffilmark{1}, S.E. Church\altaffilmark{1}, P.A.R. Ade\altaffilmark{2}, J.J.
Bock\altaffilmark{3,4}, K.M. Ganga\altaffilmark{5}, J.R.
Hinderks\altaffilmark{1}, P.D. Mauskopf\altaffilmark{2}, B.
Philhour\altaffilmark{1,3}, M.C. Runyan\altaffilmark{3,6}, K.L.
Thompson\altaffilmark{1}} \altaffiltext{1}{Stanford University,
382 Via Pueblo, Varian Building, Stanford, CA 94305}
\email{bbenson@stanford.edu} \altaffiltext{2}{Department of
Physics and Astronomy, University of Wales, Cardiff, 5, The
Parade, P.O. Box 913, Cardiff, CF24 3YB, Wales, UK}
\altaffiltext{3}{California Institute of Technology, Observational
Cosmology, M.S. 59--33, Pasadena, CA 91125} \altaffiltext{4}{Jet
Propulsion Laboratory, 4800 Oak Grove Dr., Pasadena, CA 91109}
\altaffiltext{5}{Infrared Processing and Analysis Center, MS
100-22, California Institute of Technology, Pasadena, CA 91125}
\altaffiltext{6}{University of Chicago, 5640 S. Ellis, LASR-132,
Chicago, IL 60637}

\newcommand{\lan}{\langle}
\newcommand{\ran}{\rangle}
\newcommand{\be}{\begin{equation}}
\newcommand{\ee}{\end{equation}}

\begin{abstract}
We have made measurements of the Sunyaev-Zel'dovich (SZ) effect in
six galaxy clusters at $z>0.2$ using the Sunyaev-Zel'dovich
Infrared Experiment (SuZIE II) in three frequency bands between
150 and 350\,GHz. Simultaneous multi-frequency measurements have
been used to distinguish between thermal and kinematic components
of the SZ effect, and to significantly reduce the effects of
variations in atmospheric emission which can otherwise dominate
the noise.  We have set limits to the peculiar velocities of each
cluster with respect to the Hubble flow, and have used the cluster
sample to set a 95\% confidence limit of $<1410$\,km\,s$^{-1}$ to
the bulk flow of the intermediate-redshift universe in the
direction of the CMB dipole. This is the first time that SZ
measurements have been used to constrain bulk flows. We show that
systematic uncertainties in peculiar velocity determinations from
the SZ effect are likely to be dominated by submillimeter point
sources and we discuss the level of this contamination.
\end{abstract}

\section{Introduction}
\label{intro}
The spectral distortion to the Cosmic Microwave Background
radiation (CMB) caused by the Compton scattering of CMB photons by
the hot gas in the potential wells of galaxy clusters, known as
the Sunyaev-Zel'dovich (SZ) effect, is now relatively
straightforward to detect and has now been measured in more than
50 sources \citep[see][for a review]{carl02}. Single-frequency
observations of the SZ effect can be used to determine the Hubble
constant \citep{myers97, swlh97a, poin99, reese00, poin01,
jones01, depetris02, reese02} and to measure the baryon fraction
in clusters \citep{myers97, grego01}.

\begin{figure}
\plotone{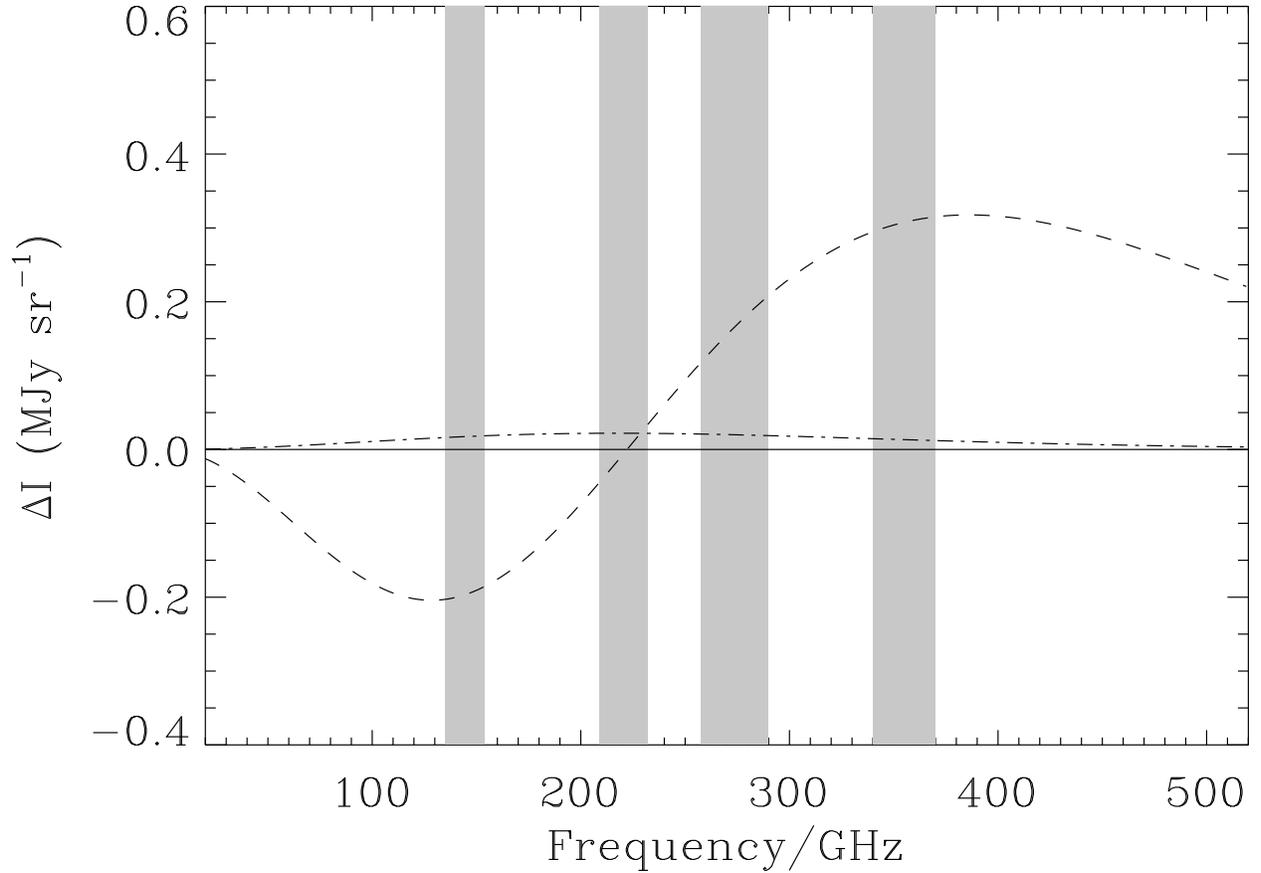} \caption[]{The frequency dependence of the SZ
effect for a cluster with optical depth $\tau=0.01$, gas
temperature 10\,keV and a peculiar velocity of $-500$\,km s$^{-1}$
(towards the observer). The thermal SZ spectrum is indicated by
the dashed line; the kinematic effect by the dot-dashed line. The
shaded regions indicate the bands in which SuZIE observes.}
\label{sec:f1}
\end{figure}

The spectrum of the SZ effect is also an important source of
information. It can be approximated by the sum of two components
(see Figure~\ref{sec:f1}) with the strongest being the thermal SZ
effect that is caused by the random thermal motions of the
scattering electrons \citep{sz1}. The kinematic SZ effect, due to
the peculiar velocity of the intracluster (IC) gas with respect to
the Hubble flow \citep{sz2}, is expected to be much weaker if
peculiar velocities are less than 1000\,km\,s$^{-1}$, as favored
by current models \citep{gramann95, sheth01, suhh02}. The thermal
SZ effect has a distinct spectral signature, appearing as a
decrement in intensity of the CMB below a frequency of $\sim$ 217
GHz, and an increment at higher frequencies (the exact frequency
at which the thermal effect is zero depends on the temperature of
the IC gas, as discussed by \citealt{yoel}). The kinematic effect
appears as a decrement at all frequencies for a cluster that is
receding with respect to the Hubble flow, and an increment at all
frequencies for a cluster that is approaching. Measurements that
span the null of the thermal effect are able to separate the two
effects, allowing the determination of the cluster peculiar
velocity \citep{swlh97b, pdm, laroque}. Additionally, SZ spectral
measurements can, in principle, be used to determine the cluster
gas temperature independently of X-ray measurements \citep{poin98,
hanse02}, the CMB temperature as a function of redshift
\citep{reph80, battis02} and also to search for populations of
non-thermal electrons \citep{shimon02}.

Peculiar velocities probe large-scale density fluctuations and
allow the distribution of matter to be determined directly without
assumptions about the relationship between light and mass.
Measurements of a large sample of peculiar velocities can be used
to probe $\Omega_m$ independently of the properties of dark energy
\citep{peel02}, and can, in principle, be used to reconstruct
modes of the gravitational potential \citep{dore02}. The local
(z$\lesssim$ 0.05) peculiar velocity field has already been
measured and has been used to place tight constraints on
$\Omega_m$ \citep{2000MNRAS.313..491B, csw, 2001astro.ph..5470C,
2001MNRAS.321..333B}. However, the techniques used cannot be
easily extended to higher redshifts because optical methods of
distance determination have errors that increase linearly with
distance. SZ spectral measurements allow peculiar velocities to be
determined independently of the extragalactic distance ladder.

In this paper we describe the measurements of the SZ spectrum of 6
galaxy clusters made with the Sunyaev-Zel'dovich Infrared
Experiment (SuZIE).  The SuZIE~II receiver makes simultaneous
measurements of the SZ effect in three frequency bands, centered
at 145\,GHz, 221\,GHz, and 355\,GHz (or 270\,GHz), spanning the
null of the SZ thermal effect. We use these measurements to set
limits on the peculiar velocity of each cluster, and to determine
the cluster optical depth. The layout of this paper is as follows:
in \S\ref{sec:sz} we define our notation for the SZ effect; in
\S\ref{sec:obs} and~\S\ref{sect:analysis} we describe the
observations and data analysis. In \S\ref{sect:errors} we consider
statistical and systematic sources of uncertainty; in
\S\ref{sect:astroconf} we estimate the level of astrophysical
confusion in our results and in \S\ref{sec:bulk_flows} we
determine the limits that the SuZIE peculiar velocity sample can
be used to set on bulk flows, and discuss future prospects for
this technique.

\section{The Sunyaev-Zel'dovich Effect}
\label{sec:sz}
\subsection{The Thermal Effect}
\label{sec:thermal}
We express the CMB intensity difference caused by a distribution
of high energy electrons, $n_e$, along the line of sight as
(Rephaeli 1995):
\be \Delta I_T = I_0 \frac{x^3}{e^x-1} [\Phi(x,T_e) - 1] \tau
\label{eqn:relthesz} \ee
where $x=h\nu/kT_0$, $I_0=2(kT_0)^2/(hc)^2$, $T_0$ is the
temperature of the CMB, $\tau=\int n_e \sigma_T dl$ is the optical
depth of the cluster to Thomson scattering, and $\Phi(x,T_e)$ is
an integral over electron velocities and scattering directions
that is specified in~\citet{yoel}. In the limit of
non-relativistic electrons, this reduces to the familiar
non-relativistic form for the thermal SZ effect:
\be [\Phi(x,T_e) - 1]  = \frac{x e^x}{e^x-1}
\left[x\frac{e^x+1}{e^x-1}-4\right]\frac{kT_e}{m_ec^2} \ee
Following Holzapfel (1997b), we define:
\be \Psi(x,T_e) = \frac{x^3}{e^x-1} [\Phi(x,T_e) - 1]
\label{eqn:psi} \ee
There also exist other analytic and numerical expressions for
equation~(\ref{eqn:relthesz}) \citep[see][]{challinor, itoh,
dolgov} based on a relativistic extension of the Kompaneets
equation \citep{komp}. These expressions are in excellent
agreement with equation~(\ref{eqn:relthesz}).

We define Comptonization as $y =  \tau \times (kT_e/m_ec^2) $. It
is a useful quantity because it represents a frequency independent
measure of the magnitude of the SZ effect in a cluster that,
unlike $\Delta I_T$, allows direct comparisons with other
experiments.

\subsection{Kinematic SZ Effect}
\label{sec:kinematic}
The change in intensity of the CMB due to the non-relativistic
kinematic SZ effect is:
\be \Delta I_K = -I_0 \times \frac{x^4 e^x}{(e^x-1)^2} \times \int
n_e \sigma_T \frac{ \textbf{v}_{p} }{c} \cdot \textbf{dl} \ee
where $\textbf{v}_{p}$ is the bulk velocity of the IC gas relative
to the CMB rest frame, and $c$ is the speed of light in units of
km\,s\,$^{-1}$. This functional form for the kinematic SZ effect
has the same spectral shape as primary CMB anisotropy
anisotropies, which represent a source of confusion to
measurements of the kinematic effect.

An analytic expression for the relativistic kinematic SZ effect
has been calculated by \cite{relkin} as a power series expansion
of $\theta_e=k T_e/mc^2$ and $\beta=v_{p}/c$, where
$v_p=\textbf{v}_p\cdot \hat{\textbf{l}}$ is the radial component
to the peculiar velocity. They found the relativistic corrections
to the intensity to be on the order of $+ 8\%$ for a cluster with
electron temperature $k T_e = 10$ keV and $v_{p}=1000$ km
s$^{-1}$. Although this is a relatively small correction to our
final results, we use their calculation in this paper.  We express
the spectral shift due to the kinematic SZ effect as:
\be \Delta I_K = -I_0 \times \tau \times \frac{ \textbf{v}_{p}
\cdot \hat{\textbf{l}} }{c} \times h(x, T_e) \ee
where $h(x, T_e)$ is given by:
\be h(x, T_e) = \frac{x^4 e^x}{(e^x-1)^2} \times \left[ 1+\theta_e
C_1(x) + \theta_e^2 C_2(x) \right] \label{eqn:hx} \ee
and $\theta_e=kT_e/m_ec^2$.  This expression includes terms up to
$O(\beta \theta_e^2$). The quantities $C_1(x)$ and $C_2(x)$ are
fully specified in \cite{relkin}, who have also calculated
corrections to equation~(\ref{eqn:hx}) up to $O(\beta^2)$.  They
find the correction from these higher order terms to be $+0.2\%$
for a cluster with $k T_e = 10$ keV and $v_{p}=1000$ km s$^{-1}$,
at a level far below the sensitivity of our observations.
Therefore it can be safely ignored.

\section{S-Z Observations}
\label{sec:obs}
\subsection{Instrument}
In this paper we report measurements of the Sunyaev-Zeldovich
effect made with the second generation Sunyaev-Zeldovich Infrared
Experiment receiver (SuZIE~II) at the Caltech Submillimeter
Observatory (CSO) located on Mauna Kea.  The SuZIE~II receiver,
described in \citet{pdm} and associated references, measures the
SZ effect simultaneously in three frequency bands.  Two of the
bands are centered at 145 and 221 GHz.  The third was originally
located at 273 GHz but has since been moved to 355 GHz to improve
the degree to which correlated atmospheric noise can be removed
from the data \citep{pdm}. Between November 1996 and November
1997, the SuZIE~II optics were changed slightly, altering both the
beam size and chop throw.  For a summary of the SuZIE~II
pass-bands and beam sizes in both configurations, see
Table~\ref{tab:passband}.  The observations discussed in this
paper include data in both configurations.

The SuZIE II instrument consists of a 2 $\times$ 2 arrangement of
3-color photometers that observe the sky simultaneously in each
frequency band.  Each frequency is detected with a 300mK NTD Ge
bolometer \citep{philbolo}.  Light is coupled to the photometers
through Winston horns which over-illuminate a 1.6K Lyot stop
placed at the image of the primary mirror formed by a warm
tertiary mirror. Each photometer defines a $\sim 1\farcm5$ FWHM
beam, with each row separated by $\sim 2\farcm3$ and each column
by $\sim 5'$ on the sky (see Figure~\ref{fig:drift}). The beam
size was chosen to correspond to typical cluster sizes at
intermediate redshift ($ 0.15 \lesssim z \lesssim 0.8$).
Bolometers in the same row that are sensitive to the same
frequency are differenced electronically to give an effective chop
throw on the sky of $5'$.  This reduces the level of common-mode
atmospheric emission as well as common mode bolometer temperature,
and amplifier gain, fluctuations. This differencing strategy has
been discussed in detail by \cite{swlh97a} and \cite{pdm}.

\subsection{Observation Strategy}
\begin{figure}
\plotone{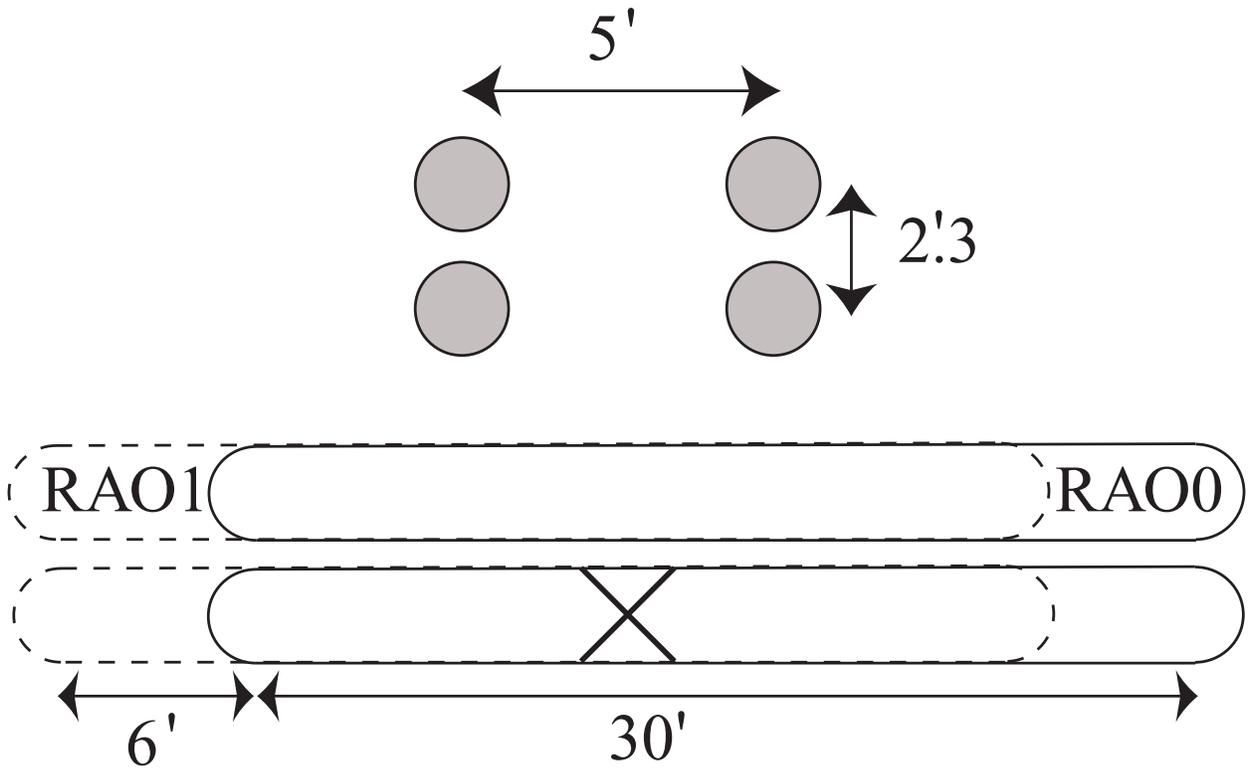} \caption[]{The upper panel shows the layout of
pixels in the SuZIE II focal plane.  The lower panel shows the
pattern traced out on the sky by a single column of the array. The
cross indicates the nominal pointing center of the observation.}
\label{fig:drift}
\end{figure}

Observations of six clusters were made with SuZIE~II over the
course of several observing runs between April 1996 and November
2000 and are summarized in Table~\ref{tab:obs}. As shown in
Figure~\ref{fig:drift}, SuZIE~II operates in a drift scanning
mode, where the telescope is pointed ahead of the source and then
parked.  The earth's rotation then causes the source to drift
through the array pixels.  Before each scan the dewar is rotated
so that the rows of the array lie along lines of constant
declination.  Each scan last two minutes, or $30'$ in right
ascension, during which time the telescope maintains a fixed
altitude and azimuth.  After a scan is complete, the telescope
reacquires the source and the scan is then repeated.  Keeping the
telescope fixed during an observation prevents slow drifts from
changes in ground-spillover from contaminating the data.  From
scan to scan the initial offset of the telescope from the source
(referred to in Figure~\ref{fig:drift} as RAO0 and RAO1) is
alternated between $12'$ and $18'$, allowing a systematic check
for an instrumental baseline and a check for any time dependant
signals. During the observations presented here, the array was
positioned so that one row passed over the center of each cluster,
as specified in Table~\ref{tab:obs}.

\subsection{The Cluster Sample}
We selected bright, known X-ray clusters from the {\em ROSAT}
X-Ray Brightest Abell Clusters \citep{1996MNRAS.281..799E,
1996MNRAS.283.1103E} and Brightest Cluster Samples
\citep{1998MNRAS.301..881E} and the Einstein Observatory Extended
Medium Sensitivity Survey \citep{1990ApJS...72..567G}. The
observations of A1835 in April 1996 were previously analyzed by
\citet{pdm}. We are using a variation of the analysis method used
by Mauskopf and choose to re-analyze this data to maintain
consistency between this cluster and the other data sets. We also
use a slightly different X-ray model for this cluster, based on a
joint analysis of radio SZ and X-ray data by \citet{reese02}. Four
of our clusters have been observed at 30\,GHz by \citet{reese02}.
A comparison of our results with their measurements will be the
subject of a separate paper.

\subsection{Calibration}
\label{sec:cal} We use Mars, Uranus and Saturn for absolute
calibration and to measure the beam shape of our instrument.  The
expected intensity of a planetary calibrator is:
\be I_{\rm plan} = \frac{\int 2k \, (\frac{\nu}{c})^2 \, T_{\rm
plan}(\nu)\,f_k(\nu)d\nu}{\int f_k({\nu})d\nu} \ee
with $T_{\rm plan}(\nu)$ being the Rayleigh-Jeans (RJ) temperature
of the planet, and $f_k(\nu)$ the transmission function of
channel~$k$, whose measurement is described later in this section.
We correct for transmission of the atmosphere by measuring the
opacity using a 225 GHz tipping tau-meter located at the CSO. This
value is converted to the opacity in each of our frequency bands
by calculating a scaling factor $\alpha_{k}$ which is measured
from sky dips during stable atmospheric conditions. For our
frequency bands at 145, 221, 273, and 355 GHz we find $\alpha =
0.8, 1.0, 2.7,$ and $5.8$.  From drift scans of the planet we
measure the voltage, $V_{\rm peak}$ that is proportional to the
intensity of the source. We then find our responsivity to a
celestial source is:
\be R = \frac{I_{\rm plan}\Omega_{\rm plan} \times e^{-\alpha
<\tau/\cos \theta_{\rm Cal}>}}{V_{peak}} \bigg[\frac{Jy}{V}\bigg]
\label{eqn:cal} \ee
where $\Omega_{\rm plan}$ is the angular size of the planet, and
$<\tau/\cos\theta_{\rm Cal}>$ is averaged over the length of the
observation, typically less than 20 minutes.  We observe at least
one calibration source every night.

The data are then calibrated by multiplying the signals by a
factor of $R \times e^{\alpha <\tau/\cos \theta_{\rm SZ}>}$. We
correct for the transmission of the atmosphere during each cluster
observation using the same method as for the calibrator
observations.  Each cluster scan is multiplied by $e^{\alpha
<\tau/\cos \theta_{\rm SZ}>}$, where $<\tau/\cos\theta_{\rm SZ}>$
is averaged over the length of that night's observation of the
cluster, typically less than three hours.  We average the
atmospheric transmission over the observation period to reduce the
noise associated with the CSO tau-meter measurement
system~\citep{2002MNRAS.336....1A}. To determine whether real
changes in $\tau$ over this time period could affect our results,
we use the maximum variation in $<\tau/\cos\theta_{\rm SZ}>$ over
a single observation, and estimate that ignoring this change
contributes a $\pm$2\% uncertainty in our overall calibration.

The uncertainty of $I_{\rm plan}$ is dominated by uncertainty in
the measurement of $T_{\rm plan}(\nu)$.  Measurements of RJ
temperatures at millimeter wavelengths exist for Uranus
\citep{griffin}, Saturn and Mars \citep{goldin}.
\citeauthor{griffin} model their measured Uranian temperature
spectrum, $T_{\rm Uranus}(\nu)$, with a third order polynomial fit
to the logarithm of wavelength. They report a 6\% uncertainty in
the brightness of Uranus.  \citeauthor{goldin} report RJ
temperatures of Mars and Saturn in four frequency bands centered
between 172 and 675 GHz. From these measurements we fit a second
order polynomial in frequency to model $T_{\rm Saturn}(\nu)$ and a
second order polynomial in the logarithm of wavelength to model
$T_{\rm Mars}(\nu)$. \citeauthor{goldin} report a $\pm$10K
uncertainty to the RJ temperature of Mars due to uncertainty from
their Martian atmospheric model, which translates to a 5\%
uncertainty in the brightness. They then use their Martian
calibration to cross-calibrate their measurements of Saturn.  Not
including their Martian calibration error, they report a $\sim$2\%
uncertainty to Saturn's RJ temperature.  Adding the Martian
calibration error in quadrature yields a total 5\% uncertainty to
the brightness of Saturn.  The rings of Saturn have an effect on
its millimeter wavelength emission which is hard to quantify. By
cross-calibrating our Saturn measurements with SuZIE~II
observations of Mars and Uranus made on the same night, we have
determined that ring angles between $\pm 9^{\circ}$ have a
negligible effect on the total emission from Saturn. At higher
ring angles we use Saturn only as a secondary calibrator. Since we
use a combination of all three planets to calibrate our data, we
estimate an overall $\pm$6\% uncertainty to $T_{\rm planet}$.

Bolometers have a responsivity that can change with the amount of
power loading on the detector; such non-linearities can
potentially affect the results of calibration on a bright planet
and the response of the detectors during the course of a night. We
have used laboratory measurements to determine the dependence of
responsivity on optical power loading.  We estimate the variation
in our loading from analysis of sky-dips taken at the telescope,
and the calculated power received from Saturn, which is the
brightest calibrator that we use.  Over this range of loading
conditions the maximum change in detector response is $\sim 7.0\%,
8.0\%,$ and $3.5\%$ in our 145, 221, and 355 GHz frequency bands
respectively. Since the responsivity change will be smaller for
the majority of our observations, we assign a 6\% uncertainty in
the calibration error budget for this effect.

\cite{pdm} has found that the SuZIE~II beam shapes have a
systematic dependence on the rotation angle of the dewar, which
affects the overall calibration of the instrument.  Based on these
measurements we assign a $\pm$5\% calibration uncertainty from
this effect. Further uncertainty to the calibration arises from
our measurement of the spectral response, $f_k(\nu)$, which
affects both the intensity that we measure from the planetary
calibrators and the SZ intensity. The spectral response of each
SuZIE~II channel was measured with a Michelson Fourier Transform
Spectrometer (FTS). We then use the scatter of the measurements of
the four bolometers that measure the same frequency to estimate
the uncertainty in the spectral calibration at that frequency.  We
estimate this uncertainty to be $<$1\% of the overall calibration.

The calibration uncertainties are summarized in
Table~\ref{tab:caltable}. Adding all of these sources in
quadrature, we estimate the total calibration uncertainty of
SuZIE~II in each of its spectral bands to be $\pm$10\%.

\subsection{Definition of the Data Set and Raw Data Processing}
\label{sec:dataset}
We now define some notation.  Each photometer contains three
bolometers each observing at a different frequency. During a scan,
the six difference signals that correspond to the spatially
chopped intensity on the sky at three frequencies, and in two
rows, are recorded.  The differenced signal is defined as:
\be D_{k} = S^{+}_{k} -S^{-}_{k} \ee
where $S_k^\pm$ is the signal from each bolometer in the
differenced pair. Because of the spacing of the photometers, this
difference corresponds to a $5'$ chop on the sky. The subscripts
$k =1, 2, 3$ refer to the frequency bands of 355 GHz (or 273 GHz),
221 GHz and 145 GHz respectively in the row that is on the source.
The subscripts $k=4, 5, 6$ refer to the same frequency set but in
the row that is off-source (see Figure~\ref{fig:drift}).  In
addition to the differenced signal, one bolometer signal from each
pair is also recorded, to allow monitoring of common-mode signals.
These six ``single-channel'' signals are referred to as $S_k$,
where $k$ is the frequency subscript previously described.  For
example, the difference and single channel at 145 GHz of the
on-source row will be referenced as $D_{3}$ and $S_{3}$.  Both the
differences and the single channels are sampled at 7 Hz.

The first step in data analysis is to remove cosmic ray spikes by
carrying out a point by point differentiation of data from a
single scan and looking for large ($>4\sigma$) deviations from the
noise.  With a knowledge of the time constant of the bolometer and
the height of the spike, we can make a conservative estimate of
how much data is contaminated and exclude that data from our
analysis.  To account for the effect of the bolometer
time-constant, which is $\le 100$ ms, we flag a region $100$ ms
$\times \ln(V_m/\sigma)$ before, and $250$ ms $\times
\ln(V_m/\sigma)$ after the spike's maximum where $V_m$ is the
height of the spike. The data are then combined into 3 second bins
each containing 21 samples and covering a region equal to
$0\farcm75 \cos \delta$ on the sky. Bins with 11 or more
contaminated samples are excluded from further analysis. Less than
1\% of the data are discarded due to cosmic ray contamination.

\section{Analysis of Calibrated Data}
\label{sect:analysis}
Each cluster data set typically comprises several hundred drift
scans, as summarized in Table~\ref{tab:obs}. Once the data have
been despiked, binned and calibrated, we need to extract the SZ
signal from the data, and at the same time, obtain an accurate
estimate of the uncertainty. For most observing conditions
atmospheric emission dominates the emission from our source. Even
under the best conditions the NEF for the 145 GHz channel, which
is least affected by the atmosphere, is rarely below $100$ mJy
s$^{1/2}$ at the signal frequencies of interest, which are $\sim
10$ mHz, while the signals we are trying to measure are $\sim 50$
mJy. The higher frequency channels are progressively worse. In
order to improve our signal/noise, we make use of our ability to
simultaneously measure the sum of the SZ signal and the atmosphere
at three different frequencies. The different temporal and
spectral behavior of the atmosphere, compared to the SZ signal of
interest, allows us to clean the data and significantly improve
the sensitivity of our measurements.

\subsection{SZ Model}
A model for the expected spatial distribution of the SZ signal in
each scan is obtained by convolving a beam map of a planetary
calibrator with the modelled opacity of the cluster.  Beam shapes
are measured by performing raster scans of a planetary calibrator
and recording the voltage response of the detectors,
$V_{k}(\theta,\phi)$ where $\theta$ is measured in the direction
of RA and $\phi$ in the direction of declination. We approximate
the electron density of the cluster with a spherically-symmetric
isothermal $\beta$ model \citep{betaref1, betaref2}:
\be n_e(r) = n_{e0} \left[ 1+\frac{r^2}{r_c^2} \right] ^{-3 \beta
/ 2} \ee
where $r$ is distance from the cluster center, and $\beta$ and
$r_c$ are parameters of the model. By integrating $n_e$ along the
line of sight, the cluster optical depth:
\be \tau(\theta, \phi) = \tau_{0} \left[
1+\frac{(\theta^2+\phi^2)}{\theta_c^2} \right] ^{(1-3 \beta) / 2}
\ee
is obtained, where linear distance $r$ has been replaced with
angles on the sky, $\theta$ and $\phi$. The model parameters of
the intra-cluster gas, $\beta$ and $\theta_c$, for each cluster
are taken from the literature and are listed in
Table~\ref{tab:icmodel} with associated references. We can now
calculate a spatial model, $m_k(\theta)$, for each cluster:
\be m_{k}(\theta) = \int \int \frac{V_{k}(\theta ',\phi
')}{V_{peak}} \frac{\tau(\theta '-\theta, \phi')}{\tau_0} d\theta
' d\phi ' \label{eqn:beamfill} \ee
that has units of steradians, and is calculated at $0'.05\times
cos \delta$ intervals for a given offset, $\theta$, in right
ascension from the cluster center.  We calculate our SZ model by
multiplying the source model by thermal and kinematic
band-averaged spectral factors given by:
\be T_k = I_0 \times \frac{m_e c^2}{k T_e} \times \frac{\int
\Psi(x,T_e) \times f_k(x) dx}{\int f_k(x) dx} \label{eqn:thermk}
\ee
and
\be K_k = -I_0 \times \frac{m_e c^2}{k T_e} \times \frac{
\mathbf{\hat{n}}_{v} \cdot \mathbf{\hat{l}} }{c}  \times
\frac{\int h(x, T_e) \times f_k(x) dx}{\int f_k(x) dx}
\label{eqn:kink} \ee
where the spectral functions $\Psi(x,T_e)$ and $h(x, T_e)$ were
previously defined in \S\ref{sec:sz}, and $f_k(x)$ is the spectral
response of channel~$k$. The vector $\mathbf{\hat{n}}_{v}$ is a
unit vector in the direction of the cluster peculiar velocity. The
quantities $T_k \times m_{k}(\theta)$ and $K_k \times
m_{k}(\theta)$ are then the SZ models for the expected responses
of frequency band $k$ to a scan across a cluster of unity central
comptonization, $y_0$, with a radial component to the peculiar
velocity, $v_{p}$, of 1\,km\,s$^{-1}$.  The calculated SZ model is
then combined into $0'.75\times \cos \delta$ bins to match the
binned SuZIE~II data, so that we define $T_k \times
m_{k}(\theta_i)$ as the thermal SZ model in channel~$k$ for the
right ascension offset $\theta$ of bin number $i$.

\label{sec:szmodel}


\subsection{Removal of Residual Atmospheric Signal}
There are two sources of residual atmospheric noise in our data,
with different temporal spectra.  The first is incomplete
subtraction of the signal that is common to each beam because of
the finite common mode rejection ratio (CMRR) of the electronic
differencing.  This effect is minimized by slightly altering the
bias, and thus the responsivity, of one of the two detectors that
form a difference. This trimming process is carried out at the
beginning of an observing campaign and is usually left unchanged
throughout the observations. The second is a fundamental
limitation introduced by the fact that the two beams being
differenced pass through slightly different columns of atmosphere;
consequently there is a percentage of atmospheric emission that
cannot be removed by differencing. While both signals originate
from the atmosphere, their temporal properties are quite different
and are accordingly removed differently in our analysis.  In what
follows we denote each frequency channel with the subscript $k$,
each scan with the subscript $j$ and each bin within a scan with a
subscript $i$.  In this way the difference and single channel
signals at 145 GHz from scan $j$ and bin $i$ are $D_{3ji}$ and
$S_{3ji}$.

The residual common mode signal from the atmosphere in the
difference channel $D_{kji}$ is modelled as proportional to the
signal from the corresponding single channel.  We define our
common mode atmospheric template, $C_{kji}$, as $C_{kji} \equiv
S_{kji}$. Because the single channels contain a small proportion
of SZ signal, there is potential to introduce a systematic error
by removing true SZ signal. However, the effect is estimated at
less than 2\% (see \S\ref{sec:cmrr}).

To model the residual differential signal from the atmosphere we
construct a linear combination of the three differential channels
in a single row which contains no thermal or kinematic SZ signal.
For the on-source row we define our differential atmospheric
template, $A_{ji}$ as:
\be A_{ji} = \alpha D_{1ji} + \gamma D_{2ji} + D_{3ji}\ee
with a similar definition for the off-source row. The coefficients
$\alpha$ and $\gamma$ are chosen to minimize the residual SZ flux
in $A_{ji}$. We describe the construction of this template in
detail in appendix \ref{sec:difftemp} and list the values of
$\alpha$ and $\gamma$ used for the on-source row observation of
each cluster in Table~\ref{tab:alphabeta}. Removing atmospheric
signal in this way significantly increases our sensitivity;
however it has the disadvantage of introducing a correlation
between different frequency channels which must be accounted for.
Also this model is dependant on the cluster parameters used, and
is subject to uncertainties in the temperature, and spatial
distribution of, the cluster gas. We quantify the uncertainty in
the final result that this produces in section \ref{sec:cmrr}.

In addition to the atmospheric signals, we also remove a slope,
$b_{kj}$, and a constant, $a_{kj}$, such that our ``cleaned''
signal is then:
\be X_{kji} = D_{kji} - a_{kj} - ib_{kj} - (e_{kj} \times C_{kji})
- (f_{kj} \times A_{ji}) \label{eqn:cleansig} \ee
Noting that since we remove a best-fit constant offset this
implies $\sum_i X_{kji} \approx 0$.

\label{sec:atmclean}

\subsection{Determination of the Cluster Location in the Scan}
\label{sect:clus_loc}
Variations in the location of the cluster center with respect to
the nominal pointing center defined in Figure~\ref{fig:drift} can
be caused by differences in the location of the X-ray and SZ
peaks, and by CSO pointing uncertainties.  The latter are expected
to be less than $10''$. To determine the true cluster location we
first co-add all of the scans for a single cluster, as described
below, then fit the data with the SZ model described above,
allowing the source position to vary.  Note we are only able to
constrain the location in right ascension; the effects of pointing
uncertainties are discussed further in~\S\ref{sect:pointing}.

Following \citet{swlh97a}, we define $X_{ki}$, the coadded signal
at each location, $i$, as:
\be X_{ki} = \frac{\sum_{j=1}^{N_s} X_{kji}/{\rm RMS}^2_{kj}}
{\sum_{j=1}^{N_s} 1/{\rm RMS}^2_{kj}} \label{eqn:coadd}\ee
where $N_s$ is the number of scans, and each scan is weighted
according to its root-mean-square (RMS) residual defined as:
\be {\rm RMS}^2_{kj} = \frac{\sum_{i=1}^{N_b}\,X^2_{kji}}{N_b-1}
\label{eq:rms_coadd}
\ee
where $N_b$ is the number of bins in a single scan.  The
uncertainty of each bin in the co-added scan, is estimated from
the dispersion about the mean value weighted by the ${\rm
RMS}_{kj}$ of each scan,
\be \sigma_{ki} =
\sqrt{\frac{\sum_{j=1}^{N_s}\,(X_{ki}-X_{kji})^2/{\rm RMS}^2_{kj}}
{(N_s-1)\sum_{j=1}^{N_s} 1/{\rm RMS}^2_{kj}}}
\label{eqn:coadduncert} \ee
This expression provides an unbiased estimate of the uncertainty
associated with each bin.

The on-source row at $\nu\sim145$\,GHz ($k=3$) provides the
highest sensitivity measurement of the cluster intensity, and so
it alone is used to fix the cluster location.  The co-added data
are fit to a model that includes an offset, $a$, a slope, $b$, and
the SZ model, where the cluster location, ${\rm RA}_{\rm offset}$
and the central comptonization, $y_0$, are allowed to vary.  For
each set of parameters, we can define $\chi^2$ as:
\be \chi^2 = \sum_{i=1}^{N_b} \frac {[X_{3i} - \{y_0 \times T_3
\times m_{3}(\theta_i-{\rm RA}_{\rm offset})\} - a -
ib]}{\sigma_{3i}^2}
\label{eqn:coaddfit}\ee
To determine the best-fit model all four parameters ($a, b, y_0, $
and the ${\rm RA}_{{\rm offset}}$) are allowed to vary while the
$\chi^2$ is minimized.  Here we are making the assumption that the
measured SZ emission in the $145$\,GHz band is entirely thermal.
We are not yet concerned with distinguishing thermal from
kinematic SZ emission because at this stage our goal is only to
fit the location of the cluster.  The cluster locations determined
in this way are listed in Table~\ref{tab:finalfit}.  All of the
clusters lie within $30''$ of the nominal pointing center, and in
most cases the cluster is located at the pointing center, within
our experimental uncertainty.

For several reasons we use the coadded method {\em only} to
determine the cluster location, not to determine $y_0$ and
$v_{p}$. These reasons include one pointed out in \citet{swlh97a},
which is that if the source contributes significantly to the
variance of each scan, then the RMS given by
equation~(\ref{eq:rms_coadd}) will be biased. This does not affect
the determination of the cluster location.  Although we could
correct this bias by subtracting the best-fit model from the data
prior to estimating the RMS, in an iterative fashion, there is
another more serious complication to the coadded data -- that of
correlations between the bins in the co-added scan produced by the
presence of residual atmospheric noise in the data, and by the
atmospheric removal process itself. Neither the bias of the RMS or
the correlations in the coadded scan affect the determination of
the cluster center, but they do need to be correctly accounted for
in the determination of the SZ parameters.

\subsection{Individual Scan Fits for Comptonization and Peculiar
Velocity}
\label{sec:sstvfits}
To generate an unbiased estimate of the SZ parameters we fit for
comptonization and peculiar velocity in all three frequency
channels simultaneously, using the cluster central position
determined from the coadded data. Following \citet{swlh97a,
swlh97b} and \citet{pdm}, we fit the data on a scan-by-scan basis
to estimate the uncertainty in the fitted parameters, because we
expect no scan-to-scan correlation in the noise. While unbiased
and producing satisfactory results, this method is not formally
optimal. An alternative would be to calculate, and then invert,
the noise covariance matrix for the data set. However, because of
the high degree of correlation in the raw data, this technique has
not been found to yield stable solutions.

We begin again with the de-spiked, binned, calibrated data defined
in section~\ref{sec:dataset}. We fit the data vector from each
scan with a slope, a constant, the model for residual common-mode
and differential atmospheric signals and an SZ model with thermal
and kinematic components.  Within each scan we allow the slope,
constant, and atmospheric coefficients to vary between frequency
channels, but we fix the comptonization and peculiar velocity to
be the same at each frequency. The residual signal left after
removal of all modelled sources of signal is then:
\be R_{kji} = D_{kji} - a_{kj} - ib_{kj} - (e_{kj} \times C_{kji})
- (f_{\nu kj} \times A_{ji}) - y_{0j} T_k m_{k}(\theta_i-{\rm
RA}_{\rm offset}) - (y_0 v_{p})_{j} K_k m_{k}(\theta_i-{\rm
RA}_{\rm offset})
\label{eqn:sstvfits} \ee
where $a_{kj}$ are the offset terms, $b_{kj}$ are the slope terms,
and $e_{kj}$ ($f_{kj}$) are the coefficients that are proportional
to the common-mode (differential-mode) atmospheric signal in
frequency channels $k=1,2,3$. The SZ-model parameters $y_{0j}$ and
$(y_0 v_{p})_{j}$ are proportional to the magnitude of the thermal
and kinematic components in each frequency channel. The
common-mode and differential atmospheric templates, $C_{kji}$ and
$A_{ji}$, are constructed using the method described in section
\ref{sec:atmclean}.  The thermal and kinematic SZ model templates,
$T_k m_{k}(\theta_i-{\rm RA}_{\rm offset})$ and $K_k
m_{k}(\theta_i-{\rm RA}_{\rm offset})$, are described in section
\ref{sec:szmodel}.

The best-fit model of scan $j$ is then determined by minimizing
the $\chi^2$, which is defined as:
\be \chi^2_{j} = \sum_{k =1}^{3} \frac{\sum_{i=1}^{N_b}
R_{kji}^2}{{\rm RMS}^2_{kj}} \label{eqn:bestfit} \ee
where
\be {\rm RMS}^2_{kj} = \frac{1}{N_b-1} \sum_i^{N_b}
(R_{kji}^{best})^2
\label{eqn:bestrms} \ee
is the mean squared of the residual signal after removal of the
best-fit model.  This has to be an iterative process because we
cannot correctly calculate the best fit model and its associated
uncertainty until we know the RMS of the residual signal with the
best-fit model removed.  As a first guess we use the RMS of the
raw data, and upon each iteration afterwards calculate the RMS
with the best-fit model removed from the previous minimization.
This process is continued until the best-fit values for $y_{0j}$
and $(y_0 v_{p})_{j}$ vary by less than one part in a million, a
condition usually met by the third iteration.  We calculate the
uncertainty of $y_{0j}$ and $(y_0 v_{p})_{j}$, which we define as
$\sigma_{y j}$ and $\sigma_{(y v_{p}) j}$, from the curvature of
$\chi^2_{j}$ in the region of the minimum.


\subsection{Likelihood Analysis of Individual Scan Fits}
\begin{figure}
\plotone{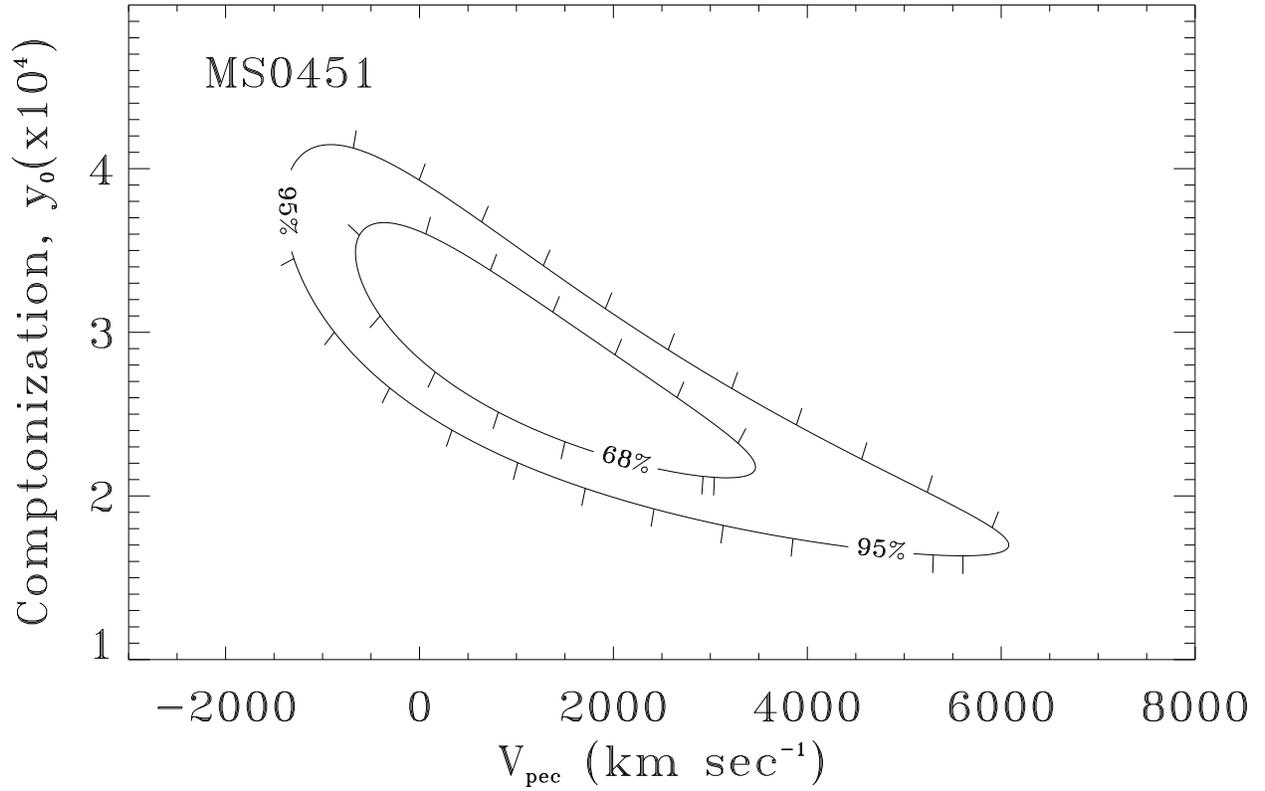} \caption[]{Results of the 2-d likelihood fit of
the combined measurements of MS0451 in November 1996, 1997, and
2000. The 68.3\% and 95.4\% confidence regions are shown for peak
Comptonization and peculiar velocity.} \label{fig:mscomblike}
\end{figure}

From the individual scan fits for comptonization and peculiar
velocity we next define a symmetric 2 by 2 covariance matrix,
$\mathbf{\Sigma}$, defined by

\be  \mathbf{\Sigma}_{11} = \frac{1}{N_s-1} \frac{ \sum_j^{N_s}
(y_{0j}-<y_0>)^2 / \sigma^4_{y j}
  } {\sum_j^{N_s} 1/\sigma^4_{y j} }
\ee
\be \mathbf{\Sigma}_{22} = \frac{1}{N_s-1} \frac{ \sum_j^{N_s}
((y_0 v_{p})_{j}-<y_0 v_{p}> )^2 / \sigma^4_{(y_0 v_{p}) j} }
{\sum_j^{N_s} 1/\sigma^4_{(y_0 v_{p}) j} } \ee
\be \mathbf{\Sigma}_{mn} = \frac{1}{N_s-1} \frac{ \sum_j^{N_s}
[(y_0 v_{p})_{j}-<y_0 v_{p}> ][y_{0j}-<y_0>] / [\sigma^2_{y j}
\sigma^2_{(y v_{p}) j}]  } {\sum_j^{N_s} 1/[\sigma^2_{y j}
\sigma^2_{(y v_{p}) j}]}  \quad \mbox{for $m \neq n$}\ee
where $y_{0j}$, $(y_0 v_{p})_j$, $\sigma_{y j}$ and $\sigma_{(y
v_{p}) j}$ are determined for scan $j$ from the minimization of
the $\chi_{j}^2$ defined in equation (\ref{eqn:bestfit}). The
quantities $<y_0>$ and $<y_0 v_{p}>$ are weighted averages of the
individual scan fits for the thermal and kinematic SZ components,
and are defined as:
\be <y_0> = \frac{\sum_j^{N_s} y_{0j}/\sigma^2_{y j}}{\sum_j^{N_s}
1/\sigma^2_{y j}} \label{eqn:ssy0} \ee \be <y_0 v_{p}> =
\frac{\sum_j^{N_s} (y_0 v_{p})_{j}/\sigma^2_{(y_0 v_{p})
j}}{\sum_j^{N_s} 1/\sigma^2_{(y v_{p}) j}} \label{eqn:ssy0vp} \ee
These weighted averages are unbiased estimators of the optical
depth and peculiar velocity.  Having calculated the covariance
matrix we define the likelihood function for our model parameters
$v_{p}$ and $y_0$ as:
\be L(v_{p},y_0) = \frac{1}{(2 \pi) |\mathbf{\Sigma}|^{1/2} } \exp
\bigg[ -\frac{1}{2} \left(
\begin{array}{c} <y_0>-y_0 \\ <y_0 v_{p}>-y_0 \times v_{p} \end{array} \right)^T \mathbf{\Sigma}^{-1}
\left(
\begin{array}{c} <y_0>-y_0 \\ <y_0 v_{p}>-y_0 \times v_{p} \end{array} \right)
 \bigg]
\ee
The likelihood is calculated over a large grid in parameter space
with a resolution of $\Delta v_{p} = 25$\,km\,s$^{-1}$ and $\Delta
y = 10^{-6}$.  As an example, the combined likelihood for the
measurements of MS0451 in November 1996, 1997, and 2000 is shown
in Figure~\ref{fig:mscomblike}.  The degeneracy between a
decreasing comptonization and an increasing peculiar velocity is a
general characteristic of the likelihood function of each cluster.
The 1-$\sigma$ uncertainty on each parameter is then determined
using the standard method of marginalizing the likelihood function
over the other parameter.  The results for each cluster are shown
in Table~\ref{tab:finalfit}.

A comparison of the numbers in Table~\ref{tab:finalfit} for A1835
with earlier results reported in \citet{pdm} that gave a
$y$-parameter of $(6.3\pm1.8)\times10^{-4}$ and a peculiar
velocity of $900\pm1500$~km\,s\,$^{-1}$ shows that the agreement
is good.

\label{sec:liketvfits}

\subsection{Spectral Plots for Each Cluster}
\begin{figure}
\plotone{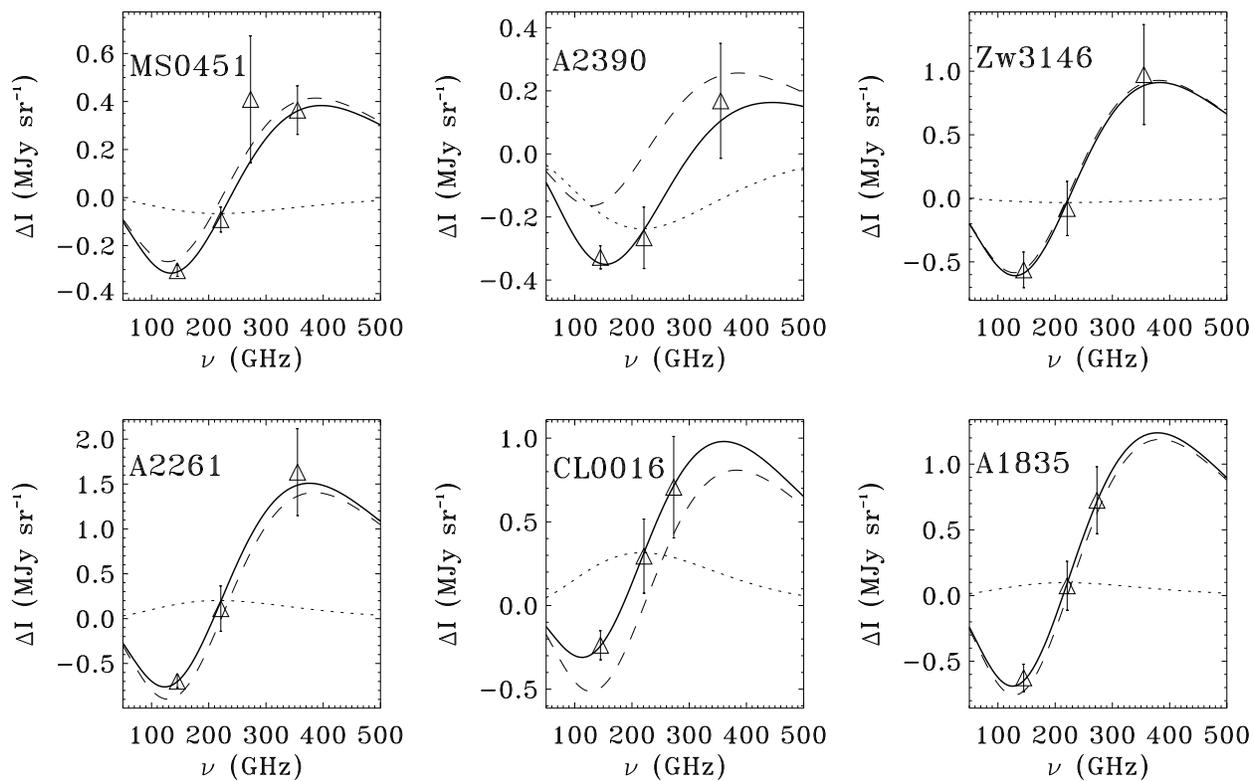} \caption[]{The measured spectrum for each cluster
in the SuZIE~II sample.  In each plot the solid line is the
best-fit SZ model, the dashed line is the thermal component of the
SZ effect and the dotted line is the kinematic component of the SZ
effect} \label{fig:clusterspectra}
\end{figure}

Figure \ref{fig:clusterspectra} plots the best-fit SZ spectrum for
each cluster with the SuZIE~II-determined intensities at each of
our three frequencies overlaid. Note, {\em these plots are for
display purposes only to verify visually that we do indeed measure
an SZ-type spectrum}. Although the values of the intensity at each
frequency are correct, the uncertainties are strongly correlated.
Consequently these intensity measurements cannot be directly
fitted to determine SZ, and other, parameters.  This is why we use
the full scan-by-scan analysis described in the previous section.

In order to calculate the points shown in
Figure~\ref{fig:clusterspectra} we calculate a new coadd of the
data at each frequency after cleaning atmospheric noise from the
data.  We define the cleaned data set, $Y_{kji}$, as:
\be Y_{kji} = D_{kji} - a_{kj} - ib_{kj} - (e_{kj} \times C_{kji})
- (f_{kj} \times A_{kji}) \label{eqn:cleanedscan} \ee
with the best-fit parameters for $a_{kj}$, $b_{kj}$, $e_{kj}$, and
$f_{kj}$ determined from equation~(\ref{eqn:bestfit}).  This
cleaned data set can now be co-added using the residual RMS
defined in equation~(\ref{eqn:bestrms}) as a weight, such that:
\be Y_{ki} = \frac{\sum_{j=1}^{N_s} Y_{kji}/{\rm RMS}^2_{kj}}
{\sum_{j=1}^{N_s} 1/{\rm RMS}^2_{kj}} \label{eqn:cleanedcoadd}\ee
Unlike equation~(\ref{eq:rms_coadd}) used in
\S\ref{sect:clus_loc}, this calculation of the RMS is not biased
by any contribution from the SZ source. The uncertainty of each
co-added bin, $\sigma_{ki}$, is determined from the dispersion
about the mean value, $Y_{ki}$, weighted by the $RMS_{kj}^2$ of
each scan,
\be \sigma_{ki} =
\sqrt{\frac{\sum_{j=1}^{N_s}\,(Y_{ki}-Y_{kji})^2/{\rm RMS}^2_{kj}}
{(N_s-1)\sum_{j=1}^{N_s} 1/{\rm RMS}^2_{kj}}} \ee

\begin{figure}
\plotone{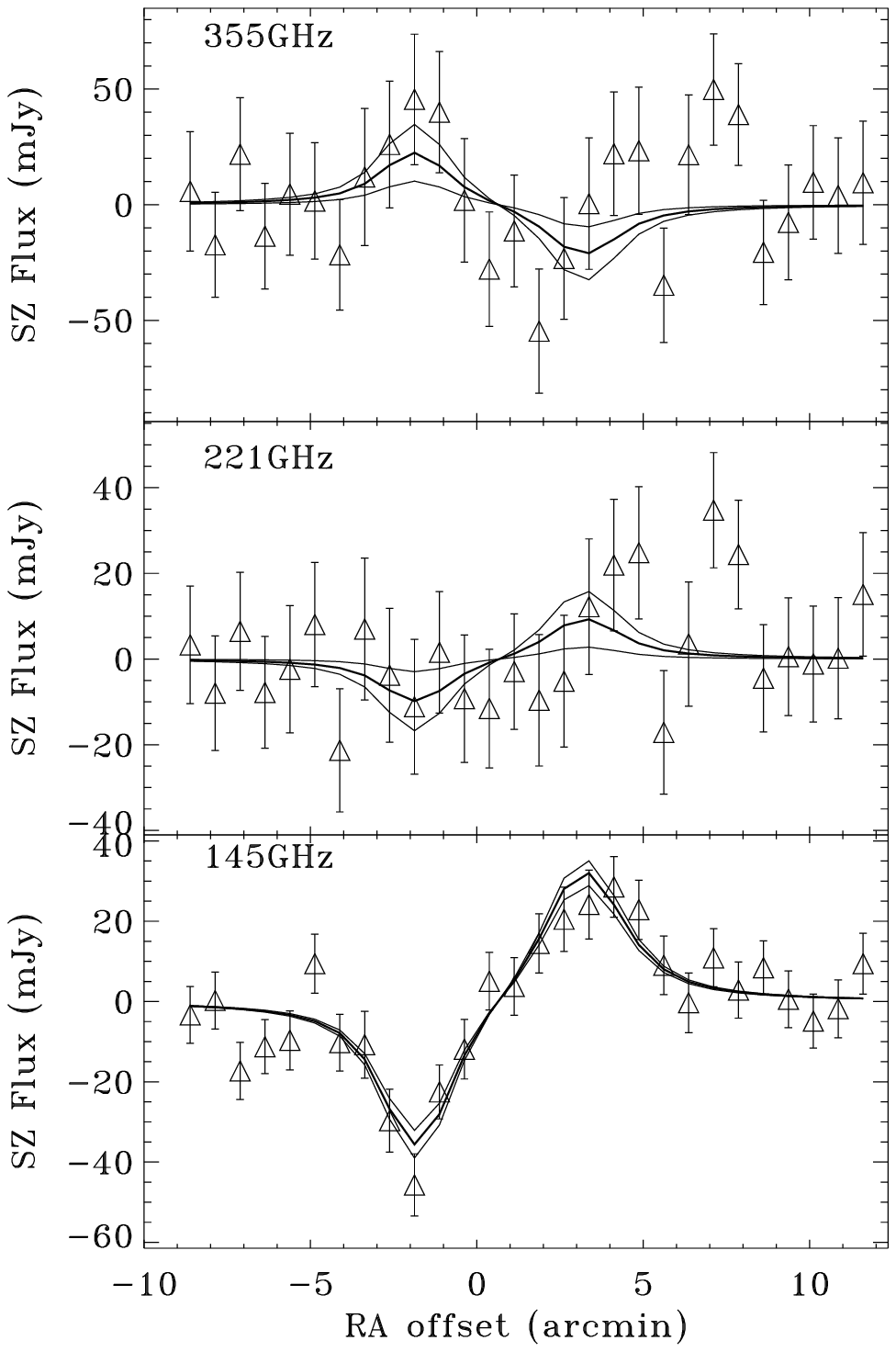} \caption[]{Co-added scans of MS0451 from November
1997 for the on-source row for each of the three SuZIE frequency
bands. The heavy line is the best-fit model to the co-added scan,
while the lighter lines represent the 1-$\sigma$ uncertainty to
those fits.} \label{fig:ms1997coadd}
\end{figure}

The best-fit central intensity, $I_k$, for each frequency band is
then found by minimizing the $\chi_k^2$ of the fit to the co-added
data, where $\chi_k^2$ is defined as follows:
\be \chi_k^2 = \sum_{i=1}^{N_b} \frac{\left[ Y_{ki}-I_k \times
m_{ki}(\theta_i-{\rm RA}_{\rm offset}) \right] }{ \sigma_{ki}^2 }
\ee
We calculate confidence intervals for $I_k$ using a maximum
likelihood estimator, $L(I_k) \propto \exp (-\chi_k^2/2)$. In
Figure~\ref{fig:ms1997coadd}, we show co-added data scans for the
November 1997 observations of MS0451 for all three on-source
frequency bands. The best fit intensity at each frequency, and the
1-$\sigma$ error bars are also shown.

\begin{figure}
\plotone{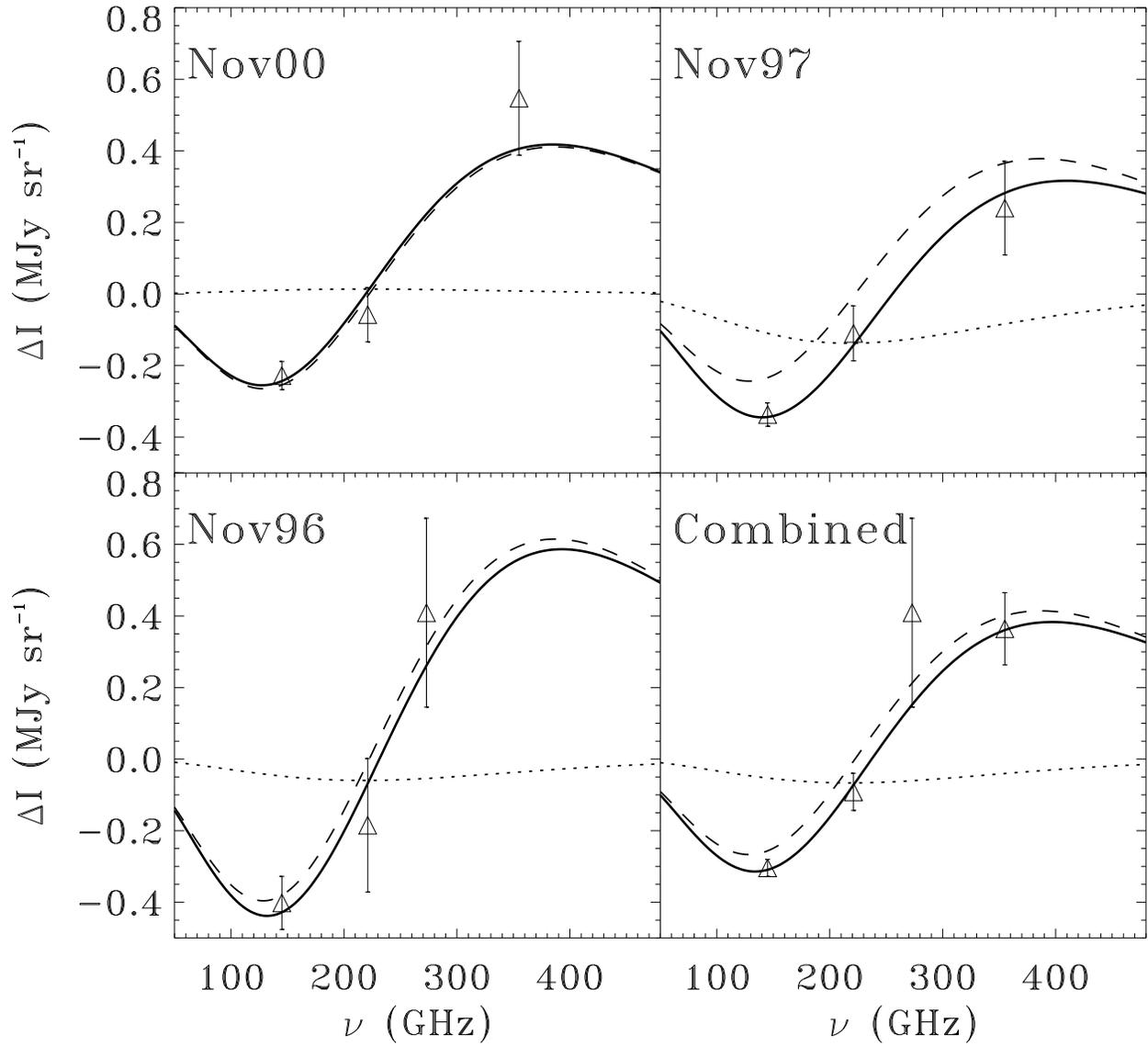} \caption[]{The measured spectra of MS0451 from
each of the three observing runs as well as the combined spectra
using the weighted mean of each spectral point. In each plot the
solid line is the best-fit SZ model, the dashed line is the
thermal component of the SZ effect and the dotted line is the
kinematic component of the SZ effect.} \label{fig:ms0451spectra}
\end{figure}

In Figure~\ref{fig:ms0451spectra}, we show the spectrum of MS0451
measured during each of the three observing runs, and the averaged
spectrum. This figure, and the best fit parameters (determined
from the scan-by-scan fitting method) shown in
Table~\ref{tab:finalfit} indicate that there is good consistency
between data sets taken many months apart.

\label{sec:ssszfits}

In order to demonstrate the value of our atmospheric subtraction
procedure, we have repeated our analysis for the MS0451 November
2000 data both with and without atmospheric subtraction.  The
derived fluxes from the coadded data are shown in
Table~\ref{tab:atm_sub}. The improvement in the sensitivity,
especially at 220 and 355\,GHz, is substantial.







\section{Additional Sources of Uncertainty}
\label{sect:errors}
The results given in Table~\ref{tab:finalfit} do not include other
potential sources of uncertainty in the data, such as calibration
errors, uncertainties in the X-ray data, and systematic effects
associated with our data acquisition and analysis techniques. We
now show that these uncertainties and systematics are negligible
compared to the statistical uncertainty associated with our SZ
measurements.  Astrophysical confusion is considered separately in
\S\ref{sect:astroconf}.

\subsection{Calibration Uncertainty}
To include the calibration uncertainty, we use a variant of the
method described in \citet{gang97}.  A flux calibration error can
be accounted for by defining a variable, $G_{k}$, such that the
correctly calibrated data is $D_{kji}^{'} = G_{k} \times D_{kji}$.
We further assume that the calibration error can be broken down
into the product of an absolute uncertainty that is common to all
frequency bands, and a relative uncertainty that differs between
frequency bands.  In this way we define $G_{k} = G^{abs} \times
G^{rel}_{k}$ with the assumption that both $G^{abs}$ and
$G^{rel}_{k}$ can be well-described by Gaussian distributions that
are centered on a value of 1.  The likelihood, marginalized over
both calibration uncertainties, is then
\be L(y_0, v_{p}) = \int^\infty_0 dG^{abs} P(G^{abs})
\int^\infty_0 L(y_0, v_{p}, G^{abs}, G^{rel}_{1}, G^{rel}_{2},
G^{rel}_{3}) \prod_{k=1}^{3} dG^{rel}_{k} P(G^{rel}_{k})
\label{eqn:callike} \ee
We evaluate these integrals by performing a 3 point Gauss-Hermite
integration \citep[see][for example]{press} using the likelihoods
calculated at the most-likely values, and the 1-$\sigma$
confidence intervals, for $G^{abs}$ and $G_k^{rel}$, which we will
now discuss. While the main source of absolute calibration error
in our data is the $\pm 6 \%$ uncertainty in the RJ temperature of
Mars and Saturn (see Section \ref{sec:cal}), it is not
straightforward to label other calibration uncertainties as either
absolute or relative. Instead we calculate equation
(\ref{eqn:callike}) assuming two different calibration scenarios:
one where our calibration uncertainty is entirely absolute such
that $G^{abs}=[0.9, 1.00, 1.1]$ and $G^{rel}_{k} = [1.0, 1.0,
1.0]$ for all $k$, and the other which has a equal combination of
the two with $G^{abs}=[0.93, 1.00, 1.07]$ and $G^{rel}_{k} =
[0.93, 1.00, 1.07]$ for all values of $k$.  This allows us to
assess whether the assignment of the error is important.

We have recalculated the best fit $y_0$ and $v_p$ using the MS0451
data taken in November 2000.  We choose this data set because it
has some of the lowest uncertainties of any of our data sets and
consequently we would expect it to be the most susceptible to
calibration uncertainties. Ignoring the calibration uncertainty,
we calculate $y_0=3.06^{+0.83}_{-0.83}\times 10^{-4}$ and
$v_{p}=-300^{+1950}_{-1250}$ km s$^{-1}$ from marginalizing the
likelihood as described in \S\ref{sec:liketvfits}. We then let
$G^{abs}$ and $G_k^{rel}$ vary over their allowed range to
calculate $L(y_0, v_{p}, G^{abs}, G_k^{rel})$ with a parameter
space resolution of $\Delta y_0 = 10^{-6}$ and $\Delta v_{p} =
25$\,km\,s$^{-1}$ in our two different calibration scenarios.
Assuming only absolute calibration uncertainty we marginalize this
likelihood over $G^{abs}=[0.9, 1.00, 1.1]$ and $G^{rel}_{k} =
[1.0, 1.0, 1.0]$ and find that $y_0=3.02^{+0.84}_{-0.83}\times
10^{-4}$ and $v_{p}=-300^{+1950}_{-1250}$ km s$^{-1}$, values
which are completely unchanged from the best fit values assuming
no calibration uncertainty. Assuming a combination of absolute and
relative calibration uncertainty we marginalize this likelihood
over $G^{abs}=[0.93, 1.00, 1.07]$ and $G^{rel}_{k} = [0.93, 1.00,
1.07]$. We find new best fit values of
$y_0=3.02^{+0.83}_{-0.84}\times 10^{-4}$ and
$v_{p}=-275^{+1950}_{-1250}$ km s$^{-1}$, again virtually
identical to the values obtained assuming no calibration
uncertainty. Therefore we conclude that for all our clusters the
error introduced from calibration uncertainty, regardless of
source, is negligible compared to the statistical error of the
measurement. The effects of calibration uncertainties are
summarized in Table~\ref{tab:cal_gas_err}.

\subsection{Gas Density and Temperature Model Uncertainties}
\label{sec:icmodelerror}

We now account for the effect of uncertainties in the $\beta$
model parameters for the intra-cluster gas by fitting our SZ data
with the allowed range of gas models based on the 1-$\sigma$
uncertainties quoted for $\beta$ and $\theta_{c}$ in
Table~\ref{tab:icmodel}.  Ideally one would fit the X-ray and SZ
data simultaneously to determine the best-fit gas model
parameters. For several of our clusters this has been done using
30\,GHz SZ maps by \cite{reese02}, and we use the values for the
$\beta$ model derived in this way. SuZIE~II lacks sufficient
spatial resolution to significantly improve on constraints from
X-ray data, and so for clusters that are not in the
\citet{reese02} sample, we use the uncertainties derived from
X-ray measurements alone.

Using a similar method to the calibration error analysis in the
previous section, we assume that the range of allowable gas models
can be well-approximated by a Gaussian distribution centered
around the most-likely value and marginalize the resulting
likelihood integrals over $\beta$ and $\theta_{core}$ individually
using 3 point Gauss-Hermite integration. In reality, the gas model
parameters $\beta$ and $\theta_{core}$ are degenerate and their
joint probability distribution is not well-approximated by two
independent Gaussians. However, this crude assumption allows us to
show below that this source of error is relatively negligible
compared to the statistical error of our results.

To estimate the effects of density model uncertainties in our
sample we study the effect on MS0451 because it has one of the
least well constrained density models from our sample.  We find
that when the allowable range of uncertainty on $\beta$ and
$\theta_c$ is included, the best fit SZ parameters are
$y_0=3.04^{+0.84}_{-0.83}\times 10^{-4}$ and
$v_{p}=-300^{+1950}_{-1250}$ km s$^{-1}$, unchanged from the
values in table~\ref{tab:finalfit}. Therefore we conclude that the
error from density model uncertainties is negligible compared to
the statistical error of the measurement.

To estimate the effects of temperature model uncertainties in our
sample we again use MS0451 because it has one of the least
constrained electron temperatures from our sample. We again assume
that the range of allowable temperatures is well approximated by a
Gaussian distribution centered around the most-likely value and
marginalize the resulting likelihood integrals over $T_e$ using 3
point Gauss-Hermite integration. We find
$y_0=3.05^{+0.82}_{-0.82}\times 10^{-4}$ and
$v_{p}=-275^{+1950}_{-1250}$ km s$^{-1}$, virtually unchanged from
the best fit values that assume no temperature uncertainty.
Therefore we conclude that the error from temperature
uncertainties is negligible compared to the statistical error of
the measurement. The effects of gas model uncertainties are
summarized in Table~\ref{tab:cal_gas_err}.


\subsection{Systematic Uncertainties}
We now consider effects that could cause systematic errors in our
estimates of $y_0$ and $v_p$.  These include instrumental baseline
drifts that could mimic an SZ source in our drifts scan, and
systematics introduced by our atmospheric subtraction technique.

\subsubsection{Baseline Drifts}
\label{sec:base} Previous observations using SuZIE~II, and the
single-frequency SuZIE~I receiver, have found no significant
instrumental baseline \citep{pdm, swlh97a, swlh97b}. Baseline
checks are performed using observations in patches of sky free of
known sources or clusters. For the data presented in this paper we
also use measurements with SuZIE~II on regions of blank sky. In
February 1998 we observed a region of blank sky at
07h40m0$^s$;$+9^{\circ}30'0''$ (J2000) for a total of $\sim$ 18
hours of integration in exceptional weather conditions. The sky
strip was $60'$ in length and was observed in exactly the same way
as the cluster observations presented in this paper. This data
represents the most sensitive measurements ever made with SuZIE~II
and consequently should be very sensitive to any residual baseline
signal (the data itself will be the subject of a separate paper).
We have repeated exactly the analysis procedure used to analyze
our cluster data with one exception, that we restrict the test
source position in the blank sky field to be within $0\farcm1$ of
the pointing center. The source positions derived from the cluster
data are consistently $\leq$ 30$''$ from the pointing center
indicating that there is no significant off-center instrumental
baseline signal. For the fit, we use a generic SZ source model
with $\beta=2/3$ and $\theta_c=20''$ and find the best-fit flux to
the co-added data in each on-source frequency band.  In our 145
GHz channel we find a best fit flux of $\Delta I = -1.05\pm 2.12$
mJy, in our 221 GHz channel we find a best fit flux of $\Delta I =
-3.56\pm 3.79$ mJy, and in our 355 GHz channel we find a best-fit
flux of $\Delta I = -1.91\pm 6.94$ mJy.  Using the November 2000
data from MS0451 as an example, if the blank sky flux measured at
145 GHz was purely thermal SZ in origin this would correspond to a
central comptonization of $y_0=0.10 \pm 0.20 \times 10^{-4}$,
while the blank sky flux measured at 221 GHz, assuming
$\tau=0.015$, corresponds to a peculiar velocity of $v_p = 635 \pm
676$\,km\,s$^{-1}$. Therefore we conclude that there is no
significant systematic due to baseline drifts in any of our three
spectral bands.


\subsubsection{Systematics Introduced by Atmospheric Subtraction }
\label{sec:cmrr}
The model that was fitted to each data set, $D_{kji}$, as defined
in equation~(\ref{eqn:sstvfits}), included common-mode atmospheric
signal, $C_{kji}$, that was defined to be proportional to the
average of our single channel signals, $S_{kji}$. While the single
channel signal is dominated by atmospheric emission variations, it
will also include some of the SZ signal we are trying to detect.
This can potentially cause us to underestimate the SZ signal in
our beam because part of it will be correlated with the single
channel template.  We estimate this effect from the correlation
coefficients $e_{kj}$ calculated during the minimization of
$\chi_j^2$ in equation (\ref{eqn:bestfit}). We estimate that the
total SZ signal subtracted out from our common-mode atmospheric
removal is $\sim 2 \%$, at a level that is negligible compared to
the statistical error of our results.

The construction of a differential atmospheric template can
potentially introduce residual SZ signal through our atmospheric
subtraction routine.  We discuss our method to construct a
differential atmospheric template in appendix \ref{sec:difftemp}
and follow the notation defined therein.  Residual SZ signal in
this template can be introduced through the simplifying assumption
that the $\alpha$ and $\gamma$ used in its construction are
spatially independent. In addition, uncertainties in the electron
temperature and density model of the cluster affect how accurately
$\alpha$ and $\gamma$ are defined.  Below we examine the effects
of these two sources of uncertainty for the November 2000
observations of MS0451.

To model the effect of a residual SZ signal in our atmospheric
template, $A_{ji}$, we re-define it by subtracting out the
expected residual thermal and kinematic signals, $Z^T_{ki}$ and
$Z^K_{ki}$ (defined in appendix \ref{sec:difftemp}), binned to
match the data set. To calculate $Z^T_{ki}$ and $Z^K_{ki}$ we use
the values of $\alpha$ and $\gamma$ given in
Table~\ref{tab:alphabeta} and assume the comptonization and
peculiar velocity values given in Table~\ref{tab:finalfit}.  For
all the clusters in our set we find $|Z_k^{T}(\theta)| < 5.7$\,mJy
and $|Z_k^{K}(\theta)| < 1.7$\,mJy across a scan of the cluster.
Using the re-defined atmospheric template we then repeat the
analysis of the data set and recalculate comptonization and
peculiar velocity. Using the November 2000 data of MS0451 as an
example, we calculate $y_0=3.06^{+0.83}_{-0.83}\times 10^{-4}$ and
$v_{p}=-300^{+1950}_{-1250}$ km s$^{-1}$ using the method
described in section \ref{sec:liketvfits}.  Using these values for
comptonization and peculiar velocity, we re-define our atmospheric
template as described above.  We then repeat our analysis routine
exactly, and calculate $y_0=3.02^{+0.82}_{-0.82}\times 10^{-4}$
and $v_{p}=-250^{+1950}_{-1275}$ km s$^{-1}$.

The accuracy of the construction of our differential atmospheric
template, parameterized by the variables $\alpha$ and $\gamma$, is
limited by our knowledge of each cluster's density model and
electron temperature.  We have calculated $\alpha$ and $\gamma$
for each cluster using the best-fit spherical beta model
parameters ($\beta, \theta_c$), and electron temperature ($T_e$).
We recalculate $\alpha$ and $\gamma$ using the $1\sigma$ range of
$T_e, \beta$ and $\theta_c$.  For the November 2000 observations
of MS0451, variations in the model parameters of the cluster cause
$\lesssim 1\%$ changes in $\alpha$ and $\gamma$.  Using the most
extreme cases of $\alpha$ and $\gamma$ we find changes of $\pm
0.01 \times 10^{-4}$ in $y_0$ and $\pm 25$ km s$^{-1}$ in $v_p$.
Adding the two sources of error of differential atmospheric
subtraction, discussed in the above paragraphs, in quadrature we
find an overall uncertainty of $^{+0.01}_{-0.04} \times 10^{-4}$
in $y_0$ and $^{+50}_{-25}$ km s$^{-1}$ in $v_p$.  We therefore
conclude that uncertainty from differential atmospheric
subtraction adds negligible error compared to the statistical
error of our results.

\subsubsection{Position Offset}
\label{sect:pointing}
In section \ref{sect:clus_loc} we allow the position of our SZ
model to vary in right ascension and determine confidence
intervals for this positional offset. However, we do not have the
necessary spatial coverage to constrain our clusters' position in
declination, $\delta$.  If a cluster's position was offset from
our pointing center in declination, we would expect the measured
peak comptonization to be underestimated from the true value. From
observations of several calibration sources over different nights,
we estimate the uncertainty in pointing SuZIE to be $\lesssim
15''$.  The cluster positions that we use are determined from
ROSAT astrometry, which is typically uncertain by $\sim 10-15''$.
Adding these uncertainties in quadrature we assign an overall
pointing uncertainty of $\Delta \delta \approxeq 20''$.

To estimate the effects of pointing uncertainty in our sample we
study the effect on observations of MS0451 in November 2000. Using
the method described in section \ref{sec:liketvfits}, which
assumed no pointing offset, we calculated
$y_0=3.06^{+0.83}_{-0.83}\times 10^{-4}$ and
$v_{p}=-300^{+1950}_{-1250}$ km s$^{-1}$.  We re-calculate the SZ
model of MS0451 with a declination offset of $20''$ from our
pointing center.  Using this SZ model we repeat our analysis
routine exactly and calculate $y_0=3.18^{+0.86}_{-0.86}\times
10^{-4}$ and $v_{p}=-300^{+1950}_{-1250}$ km s$^{-1}$.  This
corresponds to a $\sim 4\%$ underestimate of the peak
comptonization, however there is no effect on peculiar velocity.

\section{The Effects of Astrophysical Confusion}
\label{sect:astroconf}
\subsection{Primary Anisotropies}

Measurements of the kinematic SZ effect are ultimately limited by
confusion from primary CMB anisotropies which are spectrally
identical to the kinematic effect in the non-relativistic limit.
\citet{laroque} have estimated the level of CMB contamination in
the SuZIE~II bands for a conventional $\Omega=1$ ($\Lambda$CDM)
cosmology using the SuZIE~II beam size, at $|\delta y_0| < 0.05
\times 10^{-4}$ and $|\delta v_{pec}| < 380$ km s$^{-1}$.  At
present this is negligible compared to our statistical
uncertainty.

\subsection{Sub-millimeter Galaxies}
\label{sec:submmgal}

Sub-millimeter galaxies are a potential source of confusion,
especially in our higher frequency channels.  All of our clusters
have been observed with SCUBA at 450 and 850 $\mu$m and sources
detected towards all of them at 850$\mu$m \citep{smail, chapman}.
Because of the extended nature of some of these sources, it is
difficult to discern which are true point sources and which are,
in fact, residual SZ emission \citep{pdm}.  This is especially
true when the source is only detected at 850$\mu$m. We assume a
worst-case -- that all of the emission is from point sources --
and examine the effects of confusion in MS0451, A1835, and A2261.
We select these clusters because the sources in MS0451 and A2261
have SCUBA fluxes typical of all of the clusters in our set, while
A1835 has the largest integrated point source flux, as measured by
SCUBA, of all our clusters.  We consider only sources with
declinations that are within $1'$ in declination of our pointing
center since these sources will have the greatest effect on our
measurements. The point sources that meet this criterion are shown
in Table~\ref{tab:submm_sources}. To model a source observation,
we use SCUBA measurements to set the expected flux at 850$\mu$m,
and assume a spectral index, $\alpha$, of $2$ or $3$ to
extrapolate to our frequency bands, where the flux in any band is
$S_{\nu} \propto \nu^{\alpha}$. In the case of A1835, where
$450$\,$\mu$m fluxes have also been measured, the spectral indices
of the detected sources range from $1.3$--2, with large
uncertainties. For this cluster we also examine the effect of an
index of 1.7.

Note that because our beam is large we cannot simply mask out the
SCUBA sources from our scans without removing an unacceptably
large quantity of data. Instead, for each cluster, the sources are
convolved with the SuZIE~II beam-map to create a model observation
at each frequency. This model is then subtracted from each scan of
the raw data, and the entire data set is re-analyzed. The results
are summarized in Table~\ref{tab:ms0451submm}. The overall effect
of sub-millimeter point sources is to increase the measured flux
at each frequency by $10-50\%$ of the point source flux at that
frequency.  The reason that the flux error is less than the true
point source flux is that spectrally the point sources are not too
different from atmospheric emission, which also rises strongly
with frequency, and so the sources are partially removed during
our atmospheric subtraction procedure. Because the residual point
source flux is the same sign at all frequencies, it is spectrally
most similar to the kinematic effect. Consequently the effect on
the final results is to slightly over-estimate the comptonization,
with the peculiar velocity biased to a more negative value by
several hundred km\,s$^{-1}$.  While the uncertainty this
introduces is currently small compared to the statistical
uncertainty of our measurements, it is a systematic that will
present problems for future more sensitive measurements and we
expect that higher resolution observations than SCUBA will be
needed in order to accurately distinguish point sources from SZ
emission. In addition the SCUBA maps cover only $2\farcm3$, and so
all of the sources that we consider here cause a systematic
velocity towards the observer. Because of the differencing and
scan strategy that we use, sources that lie outside the SCUBA
field of view can cause an apparent peculiar velocity away from
the observer which is not quantified in this analysis.  We
conservatively estimate this contribution to be equal in magnitude
but opposite in sign to the effect of sources within the field of
view.  In reality, the effects will likely cancel to some degree,
reducing the overall uncertainty associated with submm sources.

\subsection{Unknown Sources of Systematics}
Finally, in order to check for other systematics in our data, we
calculate the average peculiar velocity of the entire set of
SuZIE~II clusters, taking into account the likelihood function of
each measurement based on the statistical uncertainties only, and
find that the average is $150^{+430}_{-390}$ km/s. This indicates
that, at present, our experimental uncertainties exceed any
systematic errors in our data.

\subsection{Summary}
Table~\ref{tab:sumuncert} summarizes the effect of all of the
known sources of uncertainty in our measurement of the peculiar
velocity of MS0451. We expect similar uncertainties for the other
clusters in our sample.  Other than the statistical uncertainty of
the measurement itself, the dominant contribution is from CMB
fluctuations and point sources.  Note that we do not include the
calculation of the baseline presented in \S\ref{sec:base} because
our measurements show no baseline at the limit set by
astrophysical confusion.




\section{Checking for Convergence of the Local Dipole Flow}
\label{sec:flow_calc}
Table~\ref{sec:t1} summarizes the current sample of SuZIE~II
measurements of peculiar velocities and includes two previous
measurements made with SuZIE~I of A2163 and A1689 \citep{swlh97b}.
Note that in each case we have assumed the statistical uncertainty
associated with each measurement, then added in quadrature an
extra uncertainty of $\pm750$\,km\,s$^{-1}$ to account for the
effects of astrophysical confusion from the CMB and submm point
sources, based on the estimates derived in \S\ref{sect:astroconf}.
These confusion estimates are larger than presented in
\citet{swlh97b}, mainly because of the more recent data on submm
sources. Consequently we use values for the peculiar velocities of
A2163 and A1689 that have uncertainties that are somewhat higher
than those previously published.

\begin{figure}
\plottwo{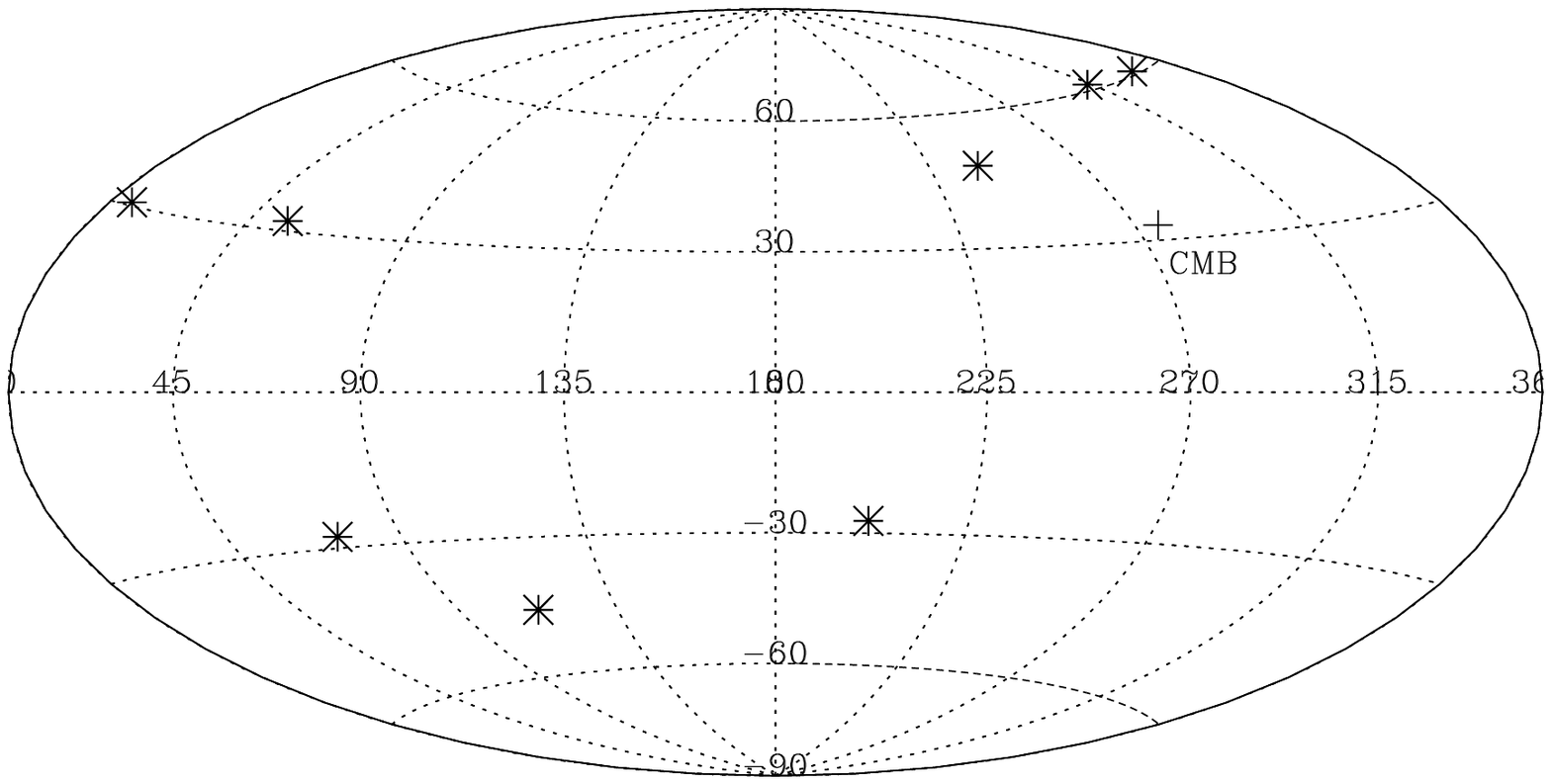}{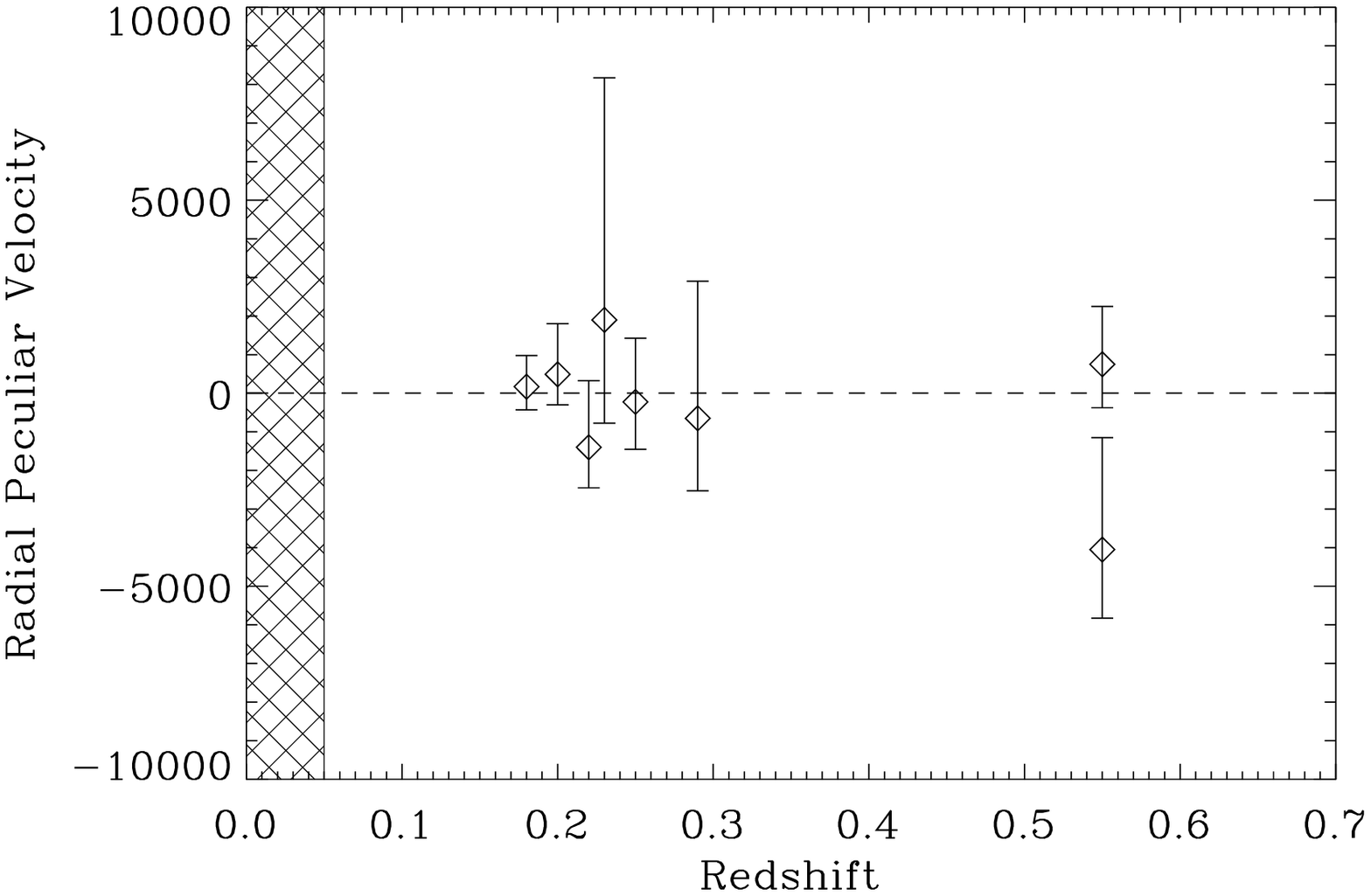} \caption[]{Left: the stars denote the
location, in galactic coordinates, of the clusters observed by
SuZIE. The direction of the CMB dipole is also shown. Right: the
measurements of the clusters plotted against redshift. The
cross-hatched region shows the range that has been probed using
optical measurements of peculiar velocities.} \label{sec:f7}
\end{figure}

The locations of the clusters on the sky are shown in
Figure~\ref{sec:f7}. The figure also shows the precision on the
radial component of each cluster peculiar velocity, plotted
against the redshift of the cluster. The cross-hatched region
shows the region of redshift space that has been probed by
existing optical surveys (see below). We now use this sample of
cluster peculiar velocities to set limits to the dipole flow at
these redshifts. The clusters in the SuZIE~II sample all lie at
distances $\gtrsim 500 h^{-1}$ Mpc, where the flow is expected to
be $\lesssim 50$ km s$^{-1}$ (see below). Of this sample, we have
taken the 6 that lie in the range $z=0.18$--0.29 and used them to
set limits on the dipole flow of structure at this redshift. This
calculation gives a useful indication of the abilities of SZ
measurements to probe large-scales motions and is also the first
time that such a measurement has been made using SZ results.

\label{sec:bulk_flows} Flows that are coherent over large regions
of space probe the longest wavelength modes of the gravitational
potential and provide a test of matter formation and evolution in
the linear region. The average bulk velocity, $V_B$, of a region
of radius, $R$, is predicted to be:
\be \left<V_B^2(R)\right>=\frac{H_0^2\Omega_m^{1.2} }{2\pi^2}
\int_0^\infty P(k) \tilde{W}^2(kR) dk \label{sect8:e2} \ee
\citep[see][for example]{willick}, where $\Omega_m$ is the
present-day matter density, $H_0$ is the Hubble constant, $P(k)$
is the matter power spectrum and $\tilde{W}(kR)$ is the Fourier
transform of the window function of the cluster sample. The shape
of the window functions depends on details of the cluster sample
such as whether all of the sphere is sampled, or whether the
clusters lie in a redshift shell \citep{1999ApJ...522..647W}. In
all models the flow is expected to converge as a function of
increasing $R$ and at $z\ge 0.18$ it should be less than
50\,km\,s$^{-1}$. Of course, equation~(\ref{sect8:e2}) is strictly
valid only in the local universe. At higher redshifts, the rate of
growth of fluctuations must be accounted for, causing the expected
bulk flow to decrease even more quickly.

All measurements of bulk flows to date have used peculiar
velocities of galaxies, determined by measuring the galaxy
redshift and comparing it to the distance determined using a
distance indicator based on galaxy luminosities, rotational
velocities, or Type Ia supernovae. These methods have yielded bulk
flows consistent with theory out to a distance of $60 h^{-1}$ Mpc
\citep[see][for reviews of the experimental
situation]{2001astro.ph..5470C}. At larger distances, the
situation is less clear.  Some null measurements do seem to
confirm that the flow converges \citep{2001MNRAS.321..277C,
2000cofl.work...25D}, but there are a number of significantly
larger flow measurements that have not yet been refuted and that
are quite discrepant with one another in direction
\citep{1999ApJ...522..647W, 1999ApJ...512L..79H,
1994ApJ...425..418L}.  The directions of these flows are shown in
Table~\ref{sec:t2}.

\begin{figure}
\plotone{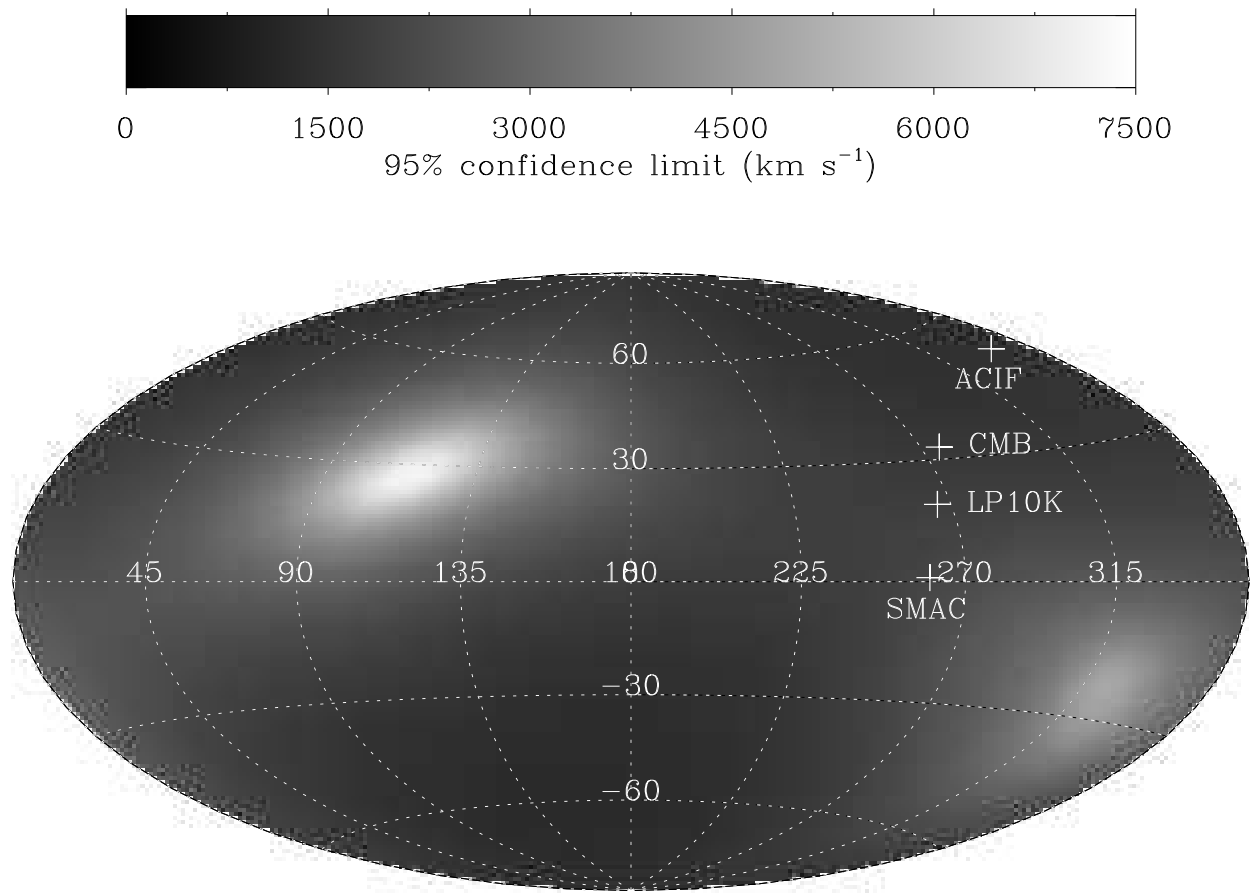} \caption[]{All-sky map showing the upper limit to
dipole flow (95\% confidence) as a function of location on the
sky. The locations of the flows listed in Table~\ref{sec:t2} are
also shown.} \label{sec:f8}
\end{figure}

We define a bulk flow as ${\bf V_{\rm B}} =(V_B, l_B, b_B)$ where
$V_B$ is the velocity of the bulk flow, $l_B$ is the galactic
longitude and $b_B$ is the galactic latitude of the flow
direction.  We can then calculate the likelihood of these
parameters as:
\be L({\bf V_{\rm B}}) = \prod\limits_{i}^{} L_i({\bf V_{\rm
B}}.\hat{r}_i) \ee
where ${\bf V_{\rm B}}.\hat{r}$ is the component of the bulk flow
in the direction of the {\em i}th cluster, and the likelihood of
the flow, given the data, $L_i$, is determined from the SuZIE~II
data.  In order to account for the effects of astrophysical
confusion, we convolve the likelihood function for the peculiar
velocity of each cluster with a gaussian probability with
$\sigma=\pm750$\,km\,s$^{-1}$ before calculating the bulk-flow
likelihood. As expected, we do not detect any bulk flow in our
data. Figure~\ref{sec:f8} shows the 95\% confidence limit to
$V_{\rm B}$ as a function of location on the sky.

Because our clusters sparsely sample the peculiar velocity field,
our upper limits are tighter in some directions than in others.
For example, we have also used our data to set limits on bulk
flows in the direction of the CMB dipole which is taken to have
coordinates $(l,b) = (276^\circ\pm 3^\circ, 33^\circ\pm 3^\circ)$
\citep{kogut94}.  We find that at 95\% confidence the flow in this
direction is $\le 1410$\,km\,s$^{-1}$. The limits towards other
directions for which optical measurements have yielded a flow
detection are shown in Table~\ref{sec:t2}.

\section{Conclusions}
We have used millimetric measurements of the SZ effect to set
limits to the peculiar velocities of six galaxy clusters. By
making measurements at three widely separated frequencies we have
been able to separate the thermal and kinematic SZ spectra.
Moreover, because our measurements in these bands are
simultaneous, we have been able to discriminate and remove
fluctuations in atmospheric emission which dominate the noise at
these wavelengths. These observations have allowed us to make the
first SZ-determined limits on bulk flows. In certain directions
the limits are approaching the level of sensitivity achieved with
optical surveys at much lower redshifts. Because our sky coverage
is not uniform, our sensitivity to bulk flows varies greatly over
the sky. We are continuing the SuZIE~II observation program with
the aim of obtaining a larger, more uniform cluster sample to
improve our bulk flow limits.

The precision with which SuZIE~II can measure peculiar velocities
is limited not only by the small number of detectors, but also by
atmospheric and instrumental noise, and astrophysical confusion
from submillimeter galaxies. In the last few years the potential
of SZ astronomy has been realized and new telescopes equipped with
bolometer arrays that will have hundreds to thousands of pixels
are planned \citep{carl02, Staggs:2001bk}. Our measurements
demonstrate the feasibility of using the SZ effect to measure
cluster peculiar velocities, but highlight some issues that need
further investigation:
\begin{enumerate}
\item The atmosphere will continue to be an issue for ground-based
measurements. We have demonstrated that wide, simultaneous
spectral coverage can significantly diminish, but not completely
remove, this source of noise.
\item Confusion from submillimeter point sources can lead to a
systematic peculiar velocity measurement. Further work is needed
to determine the best strategy for identifying and removing this
contaminant, especially for experiments that will map a large area
of sky with lower angular resolution than SuZIE~II, such as the
High Frequency Instrument (HFI) that will operate at frequencies
of 100--850 GHz as part of the Planck
satellite~\citep{2000ApL&C..37..161L}. Although high-resolution
measurements of clusters at millimeter wavelengths will be
possible with future instruments such as ALMA, it may not be
feasible to use this method to subtract astrophysical
contamination from a sample of the size needed (of order 1000+
velocities) to effectively probe large-scale structure with
SZ-determined peculiar velocities. Instruments such as SuZIE~II
will be invaluable for investigating removal techniques based on
spectral rather than spatial information, especially if more
frequency bands are incorporated into the instrument.
\end{enumerate}

The SuZIE program is supported by a National Science Foundation
grant AST-9970797 and by a Stanford Terman Fellowship awarded to
SEC. JH acknowledges support from a National Science Foundation
Graduate Research Fellowship and a Stanford Graduate Fellowship.
This work was partially carried out at the Infrared Processing and
Analysis Center and the Jet Propulsion Laboratory of the
California Institute of Technology, under a contract with the
National Aeronautics and Space Administration. We would like to
thank the CSO staff for their assistance with SuZIE observations.
The CSO is operated by the Caltech Submillimeter Observatory under
contract from the National Science Foundation.

\begin{appendix}

\section{Creating an Atmospheric Template}

SuZIE makes simultaneous multi-frequency observations of a source.
Besides the obvious advantage of instantaneously measuring the
spectrum of the SZ source, this method permits atmospheric noise
removal through spectral discrimination of the atmospheric signal.
We realize this by creating an atmospheric template with no
residual SZ signal for each row of photometers by forming a linear
combination of the three differential frequency channels in that
row on a scan by scan basis.  We define this atmospheric template,
$A_{j}(\theta)$, as:
\be A_{j}(\theta) = \alpha D_{1j}(\theta) + \gamma D_{2j}(\theta)
+ D_{3j}(\theta) \ee
where the coefficients $\alpha$ and $\gamma,$ are chosen to
minimize the residual SZ signal in $A_{j}(\theta)$.  If we define
$F_{k}^{SZ}(\theta)$ as the SZ signal in channel $D_{k}(\theta)$
this implies that the residual SZ signal in our atmospheric
template, which we define as $Z_{k}^{SZ}(\theta)$, is then
\be Z_{k}^{SZ}(\theta) = \alpha F_{1}^{SZ}(\theta) + \gamma
F_{2}^{SZ}(\theta) + F_{3}^{SZ}(\theta) \label{eqn:residualsz}\ee
where the SZ flux in each channel includes contributions for both
the thermal and kinematic effects, such that, $F_{k}^{SZ}(\theta)
= F_{k}^{SZ,T}(\theta) + F_{k}^{SZ,K}(\theta)$.  The thermal SZ
flux in each channel, using the notation of \S\ref{sec:szmodel},
is more precisely:
\be F_{k}^{SZ,T}(\theta) = y_0 \times T_k(T_e) \times
m_{k}(\theta) \ee
with $y_0$ being the central comptonization of the cluster,
$T_k(T_e)$ being defined in equation~(\ref{eqn:thermk}), and
$m_k(\theta)$ defined in equation~(\ref{eqn:beamfill}). Similarly,
the kinematic SZ flux in each channel is
\be F_{k}^{SZ,K}(\theta) = y_0 v_{p} \times K_k(T_e) \times
m_{k}(\theta) \ee
with $v_{p}$ being the peculiar velocity of the cluster, and
$K_k(T_e)$ defined in equation~(\ref{eqn:kink}).

Using this notation we rewrite our expression for the residual SZ
signal in the atmospheric template as
\be Z_{k}^{SZ}(\theta) = Z^T_{k}(\theta) + Z^K_{k}(\theta)
\label{eqn:ressz} \ee
with:
\begin{eqnarray}
Z_{k}^{T}(\theta) & = & y_0 \left[ \alpha T_1(T_e) m_1(\theta) +
\gamma T_2(T_e) m_2(\theta) + T_3(T_e) m_3(\theta) \right]
\label{eqn:resszt} \\
Z_{k}^{K}(\theta) & = & - y_0 v_p \left[ \alpha K_1(T_e)
m_1(\theta) + \gamma K_2(T_e) m_2(\theta) + K_3(T_e) m_3(\theta)
\right]
\label{eqn:resszk}
\end{eqnarray}
Solving for $Z_{k}^{SZ}(\theta) = 0$ requires a position-dependent
solution for $\alpha$ and $\gamma$ because $m_k(\theta)$ is not
constant between frequency bands. For simplicity, we use the peak
values of $m_k(\theta)$ to calculate $\alpha$ and $\gamma$ for
$Z_{k}^{SZ} = 0$.  The error introduced from this assumption is
discussed in Section \ref{sec:cmrr}.  As an aside, this template
should also be free of primary CMB anisotropy, on spatial scales
similar to our clusters, because the kinematic SZ effect is
spectrally indistinguishable from a primary CMB anisotropy.

\label{sec:difftemp}

\end{appendix}

\newpage

\begin{deluxetable}{lcccccccc}
\tablecaption{SuZIE Frequencies and Beam
Sizes\label{tab:passband}} \tablewidth{0pt}
  \tablecolumns{9}
  \tablehead{
& \multicolumn{4}{c}{1996} &
 \multicolumn{4}{c}{1997-present} \\ \cline{2-5}\cline{6-9}
 & \colhead{$\nu_0$} & \colhead{$\Delta\nu$} & \colhead{FWHM} & \colhead{$\Omega$}
  & \colhead{$\nu_0$} & \colhead{$\Delta\nu$} & \colhead{FWHM} & \colhead{$\Omega$} \\
\colhead{Channel\tablenotemark{a}} & \colhead{[GHz]} &
\colhead{[GHz]} & \colhead{arcmin} & \colhead{arcmin$^2$} &
\colhead{[GHz]} & \colhead{[GHz]} & \colhead{arcmin} &
\colhead{arcmin$^2$}
   }
    \startdata
 270/350 GHz & \nodata & \nodata & \nodata & \nodata & \nodata & \nodata & \nodata & \nodata \\
$S^+_{10}$ & 274.0 & 32.6 & 1.35 & 1.85 & 355.1 & 30.3 & 1.50 & 2.17 \\
$S^-_{10}$ & 272.8 & 28.8 & 1.30 & 1.79 & 354.2 & 31.1 & 1.54 & 2.42 \\
$S^+_{11}$ & 272.9 & 30.3 & 1.25 & 1.65 & 356.7 & 30.8 & 1.57 & 2.24 \\
$S^-_{11}$ & 272.3 & 30.6 & 1.30 & 1.76 & 354.5 & 31.5 & 1.60 & 2.58 \\
 221 GHz & \nodata & \nodata & \nodata & \nodata & \nodata & \nodata & \nodata & \nodata \\
$S^+_{20}$ & 222.2 & 22.4 & 1.30 & 1.64 & 221.4 & 21.8 & 1.36 & 1.90 \\
$S^-_{20}$ & 220.0 & 23.1 & 1.25 & 1.67 & 220.5 & 23.8 & 1.40 & 2.08 \\
$S^+_{21}$ & 220.7 & 24.8 & 1.15 & 1.53 & 221.7 & 22.9 & 1.47 & 2.02 \\
$S^-_{21}$ & 220.0 & 24.1 & 1.20 & 1.54 & 220.8 & 21.7 & 1.40 & 2.18 \\
 145 GHz & \nodata & \nodata & \nodata & \nodata & \nodata & \nodata & \nodata & \nodata \\
$S^+_{30}$ & 146.1 & 19.9 & 1.45 & 2.14 & 145.2 & 18.1 & 1.50 & 2.35 \\
$S^-_{30}$ & 146.4 & 21.3 & 1.40 & 2.16 & 144.6 & 18.6 & 1.64 & 2.73 \\
$S^+_{31}$ & 145.0 & 20.7 & 1.30 & 1.86 & 145.5 & 18.2 & 1.60 & 2.51 \\
$S^-_{31}$ & 144.4 & 19.5 & 1.35 & 1.91 & 144.9 & 17.4 & 1.60 & 2.73 \\
  \enddata
  \tablenotetext{a}{The notation is of the form $S_{{\rm freq},{\rm
  row}}$ and the $\pm$ sign refers to the sign of the channel in
  the differenced data.}
\end{deluxetable}

\clearpage


\newpage

\begin{deluxetable}{lcccccccc}
\tablecaption{Summary of SuZIE observations\label{tab:obs}}
\tablewidth{0pt}
  \tablehead{
& & \colhead{R.A.\tablenotemark{a}} & \colhead{Decl.\tablenotemark{a}} & & \colhead{Total} & \colhead{Accepted} & \colhead{Integration} \\
\colhead{Source} & \colhead{z} & \colhead{(J2000)} &
\colhead{(J2000)} & \colhead{Date} & \colhead{Scans} &
\colhead{Scans} & \colhead{Time (hours)} & \colhead{Ref.}
    }
 \startdata
    A2261  & 0.22 & 17 22 27.6 & $+$32 07 37.1 & Mar 99 & 144 & 133 & 4.4 & 1\\
    A2390 & 0.23 & 21 53 36.7 & $+$17 41 43.7 & Nov 00 & 147 & 128 & 4.3 & 1\\
    ZW3146 & 0.29 & 10 23 38.8 & $+$04 11 20.4 & Nov 00 & 151 & 131 & 4.4 &1 \\
    A1835 & 0.25 & 14 01 02.2 & $+$02 52 43.0 & Apr 96 & 638 & 577 & 19.2 & 1\\
    Cl0016$+$16 & 0.55 & 00 21 08.5 & $+$16 43 02.4 & Nov 96 & 304 & 277 & 9.2 & 2\\
                    &      &            &                & & &
& &\\
   MS0451.6$-$0305 & 0.55 & 04 54 10.8 & $-$03 00 56.8  & Nov 96 & 405 & 348 & 11.6 & 2\\
 \multicolumn{1}{c}{$''$} & & & & Nov 97 & 236 & 211 & 7.0 & \\
 \multicolumn{1}{c}{$''$} & & & & Nov 00 & 375 & 306 & 10.2  & \\
    MS0451.6$-$0305 & & & & Total & 1016 & 865 & 28.8\\
\enddata
 \tablerefs{(1) \citet{1998MNRAS.301..881E}; (2) \citet{1994ApJS...94..583G}}
 \tablenotetext{a}{Units of RA are
hours, minutes and seconds and units of declination are degrees,
arcminutes and arcseconds}
\end{deluxetable}

\begin{deluxetable}{rcccc}
\tablecaption{Break down of the calibration uncertainties
\label{tab:caltable}} \tablewidth{0pt}
  \tablehead{
    \colhead{Source} & \colhead{Uncertainty (\%)}
    }
\startdata
  Detector non-linearities & 6 \\
  Planetary temperature   & 6 \\
  Atmospheric $\tau$ & 2 \\
  Spectral response  & 1 \\
  Beam uncertainties & 5 \\
  & \\
  Total        & 10 \\
\enddata
\end{deluxetable}

\begin{deluxetable}{lcccl}
\tablecaption{IC gas temperatures beta model parameters
\label{tab:icmodel}}\tablewidth{0pt}
\tablehead{
  & \colhead{$kT_e$} & & \colhead{$\theta_{c}$} & \\
  \colhead{Cluster} & \colhead{(keV)} & \colhead{$\beta$} &
  \colhead{(arcsec)} & \colhead{Ref.}}
 \startdata
  A2261 &  $8.82^{+0.37}_{-0.32}$ & $0.516^{+0.014}_{-0.013}$ & $15.7^{+1.2}_{-1.1}$ & 1;2;2\\
  A2390 &  $10.13^{+1.22}_{-0.99}$ & $0.67^{+0.0}_{-0.0}$ & $52.0^{+0.0}_{-0.0}$ & 1;3;3 \\
  Zw3146 &  $6.41^{+0.26}_{-0.25}$ & $0.74^{+0.0}_{-0.0}$ & $13.0^{+0.0}_{-0.0}$ & 1;3;3 \\
  A1835 & $8.21^{0.19}_{-0.17}$ & $0.595^{+0.007}_{-0.005}$ & $12.2^{+0.6}_{-0.5}$ & 1;2;2 \\
  Cl0016 & $7.55^{+0.72}_{-0.58}$ & $0.749^{+0.024}_{-0.018}$ & $42.3^{+2.4}_{-2.0}$ & 4;2;2 \\
  MS0451 &  $10.4^{+1.0}_{-0.8}$ & $0.806^{+0.052}_{-0.043}$ & $34.7^{+3.9}_{-3.5}$ & 5;2;2 \\
 \enddata
  \tablerefs{(1) \cite{af98a}, (2) \cite{reese02}, (3) \cite{ettori},
  (4) \cite{hb98}, (5) \cite{donahue}}
\end{deluxetable}

\begin{deluxetable}{lcc}
\tablecaption{Differential Atmospheric Template Factors
\label{tab:alphabeta}}\tablewidth{0pt}
 \tablehead{
 \colhead{Cluster} & \colhead{$\alpha$} & \colhead{$\gamma$} 
  }
 \startdata
    A2261\tablenotemark{a} & 0.6366 & -1.4490 \\ 
    A2390\tablenotemark{a} & 0.6563 & -1.4721 \\ 
    Zw3146\tablenotemark{a} & 0.6428 & -1.3808 \\ 
    MS0451\tablenotemark{a} (Nov97) & 0.6770 & -1.3933 \\ 
    MS0451\tablenotemark{a} (Nov00) & 0.6449 & -1.4286 \\ 
    \nodata & \nodata & \nodata \\
    A1835\tablenotemark{b} & 1.2133 & -2.1745 \\ 
    Cl0016\tablenotemark{b} & 1.2353 & -2.2331 \\ 
    MS0451\tablenotemark{b} (Nov96) & 1.2246 & -2.2076 \\ 
\enddata
\tablenotetext{a}{High frequency channel was 355 GHz}
\tablenotetext{b}{High frequency channel was 273 GHz}
\end{deluxetable}

\begin{deluxetable}{lcccc}
\tablecaption{Summary of
Results\label{tab:finalfit}}\tablewidth{0pt}
 \tablehead{
   \colhead{Cluster} & \colhead{Date} & \colhead{$\Delta$ RA (arcsec)} &
   \colhead{$y_0\times 10^4$} & \colhead{$v_{pec}$ (km s$^{-1}$)}
}
 \startdata
 A2261 & Mar99 & $1.3^{+11.0}_{-11.0}$ & $7.58^{+2.24}_{-2.28} $ & $-1400^{+1725}_{-1050}$ \\
 A2390 & Nov00 & $-2.5^{+14.3}_{-14.3}$ & $1.67^{+1.03}_{-0.72} $ & $+1950^{+6275}_{-2675}$\\
 Zw3146 & Nov00 & $13.9^{+23.9}_{-22.9}$ & $3.60^{+1.79}_{-2.61} $ & $-650^{+3550}_{-1875}$ \\
 A1835 & Apr96 & $-22.5^{+12.0}_{-13.0}$ & $7.54^{+1.61}_{-1.61} $ & $-225^{+1650}_{-1225}$ \\
 Cl0016 & Nov96 & $-2.9^{+25.0}_{-26.9}$ & $2.98^{+1.32}_{-2.59} $ & $-4050^{+2900}_{-1775}$ \\
 & & & & \\
 MS0451 & Nov96 & $-1.5^{+16.0}_{-16.0}$ & $2.09^{+1.87}_{-0.88} $ & $+350^{+6625}_{-2925}$ \\
  \nodata & Nov97 & $22.0^{+9.0}_{-9.0}$ & $2.15^{+0.72}_{-0.73} $ & $+1650^{+3775}_{-2125}$\\
  \nodata& Nov00 & $-25.0^{+17.0}_{-17.0}$ & $3.06^{+0.83}_{-0.83} $ & $-300^{+1950}_{-1250}$ \\
  \nodata & Combined Fit & \nodata & $2.80^{+0.51}_{-0.52} $ & $+750^{+1500}_{-1125}$\\
 \enddata
\end{deluxetable}

\begin{deluxetable}{ccc}
\tablecaption{The Effects of Atmospheric Subtraction on Derived
Fluxes for MS0451 Nov 2000 Data
\label{tab:atm_sub}} \tablewidth{0pt}
  \tablehead{
    \colhead{Frequency} & \multicolumn{2}{c}{Flux (MJy sr$^{-1}$)} \\
    \colhead{(GHz)} & \colhead{With Atm. Subtraction} & \colhead{Without Atm. Subtraction}
    }
\startdata
  355 & $0.547^{+0.159}_{-0.160}$  & $-1.320^{+1.083}_{-1.084}$ \\
  221 & $-0.055^{+0.077}_{-0.076}$ & $-0.382^{+0.161}_{-0.162}$ \\
  145 & $-0.228^{+0.040}_{-0.039}$ & $-0.227^{+0.060}_{-0.060}$ \\
 \enddata
\end{deluxetable}

\begin{deluxetable}{lcc}
\tablecaption{Effects of Calibration and IC Gas Model
Uncertainties on MS0451 (Nov 2000)\label{tab:cal_gas_err}}
 \tablewidth{0pt}
 \tablehead{
 \colhead{Uncertainty} & \colhead{$y_0 \times 10^{4}$} & \colhead{$v_p$ (km
 s$^{-1}$)}
 }
 \startdata
 \phm{space} Statistical Uncertainty
            & $3.06^{+0.83}_{-0.83}$ & $-300^{+1950}_{-1250}$ \\
 \phm{space} Calibration\tablenotemark{a} (Absolute Only)
            & $3.02^{+0.84}_{-0.83}$ & $-300^{+1950}_{-1250}$ \\
 \phm{space} Calibration\tablenotemark{a} (Equal Absolute and Relative)
            & $3.02^{+0.83}_{-0.84}$ & $-275^{+1950}_{-1250}$ \\
 \phm{space} IC Density Model
            & $3.04^{+0.84}_{-0.83}$ & $-300^{+1950}_{-1250}$ \\
 \phm{space} IC Gas Temperature
            & $3.05^{+0.82}_{-0.82}$ & $-275^{+1950}_{-1250}$ \\
\enddata
\tablenotetext{a}{See text for details}
\end{deluxetable}

\begin{deluxetable}{rcccccc}
\tablecaption{SCUBA sources towards MS0451, A1835 and A2261.
\label{tab:submm_sources}} \tablewidth{0pt}
  \tablehead{
    & & \multicolumn{2}{c}{Source Coordinates} &\multicolumn{2}{c}{Flux (mJy)}   &  \\
    \cline{3-6}
    \colhead{Cluster} & \colhead{Source} & \colhead{RA\tablenotemark{a} (J2000)}
    & \colhead{Dec\tablenotemark{a} (J2000)} & 850$\mu$m & 450$\mu$m   & Ref.
    }
\startdata
  A2261  & SMMJ$17223+3207$ & 17 22 20.8 & +32 07 04 & $17.6\pm3.9$ & -- & 1\\
\nodata  & \nodata  &\nodata  &\nodata  &\nodata  &\nodata  & \\
  A1835  & SMMJ$14009+0252$ & 14 00 57.7 & +02 52 50 & $14.5\pm1.7$ & $33\pm7$  & 2\\
         & SMMJ$14011+0252$ & 14 01 05.0 & +02 52 25 & $12.3\pm1.7$ & $42\pm7$  & 2\\
         & SMMJ$14010+0252$ & 14 01 02.3 & +02 52 40 &  $5.4\pm1.7$ & $20\pm7$  & 2\\
\nodata  & \nodata  &\nodata  &\nodata  &\nodata  &\nodata  & \\
  MS0451 & SMMJ$04542-0301$ & 04 54 12.5 & $-$03 01 04 & $19.1\pm4.2$ & -- & 1\\
\enddata
\tablerefs{(1) \citet{chapman}; (2) \citet{smail}}
\tablenotetext{a}{Units of RA are hours, minutes and seconds and
units of declination are degrees, arcminutes and arcseconds}
\end{deluxetable}

\begin{deluxetable}{lccccccc}
\tablecaption{MS0451, A2261, and A1835 Sub-mm Galaxy
Confusion\label{tab:ms0451submm}}
 \tablewidth{0pt}
 \tablehead{
  & \colhead{Spectral} & \multicolumn{4}{c}{Flux/mJys} & & \colhead{$v_{p}$}
  \\\cline{3-6}
\colhead{Cluster} & \colhead{Model} & \colhead{145GHz} &
\colhead{221GHz} & \colhead{273GHz} & \colhead{355GHz} & $y_0
\times 10^4$ & (km s$^{-1}$)
 }
 \startdata
 MS0451 & No Source\tablenotemark{a}
     & $-24.2^{+4.1}_{-4.1}$ & $-5.4^{+6.7}_{-6.7}$ & \nodata & $+55.4^{+15.8}_{-15.7}$
     & $3.06^{+0.83}_{-0.83}$ & $-300^{+1950}_{-1250}$ \\
  & $\alpha=3$
     & $-24.9^{+4.1}_{-4.1}$ & $-7.5^{+6.7}_{-6.7}$ & \nodata & $+51.7^{+15.7}_{-15.8}$
     & $2.92^{+0.81}_{-0.83}$ & $+25^{+2150}_{-1350}$ \\
  & $\alpha=2$
     & $-26.3^{+4.1}_{-4.1}$ & $-9.5^{+6.7}_{-6.7}$ & \nodata & $+49.4^{+15.8}_{-15.7}$
     & $2.87^{+0.81}_{-0.83}$ & $+350^{+2275}_{-1425}$ \\
    \nodata  & \nodata  &\nodata  &\nodata  &\nodata  &\nodata &\nodata & \\
 A2261 & No Source\tablenotemark{a}
     & $-44.5^{+4.4}_{-4.4}$ & $+11.0^{+14.1}_{-14.1}$ & \nodata & $+113.9^{+29.7}_{-29.8}$
     & $7.58^{+2.24}_{-2.27}$ & $-1400^{+1725}_{-1050}$ \\
  & $\alpha=3$
     & $-45.0^{+4.4}_{-4.4}$ & $+8.9^{+14.1}_{-14.1}$ & \nodata & $+110.0^{+29.8}_{-29.7}$
     & $7.37^{+2.20}_{-2.31}$ & $-1250^{+1850}_{-1100}$ \\
  & $\alpha=2$
     & $-45.9^{+4.4}_{-4.4}$ & $+7.9^{+14.1}_{-14.1}$ & \nodata & $+108.9^{+29.8}_{-29.6}$
     & $7.34^{+2.20}_{-2.31}$ & $-1150^{+1850}_{-1150}$ \\
    \nodata  & \nodata  &\nodata  &\nodata  &\nodata  &\nodata &\nodata & \\
 A1835 & No Source\tablenotemark{a}
     & $-36.3^{+5.6}_{-5.7}$ & $+2.9^{+8.6}_{-8.6}$ & $+38.2^{+12.6}_{-12.7}$
     &\nodata
     & $7.54^{+1.60}_{-1.61}$ & $-200^{+1650}_{-1250}$ \\
  & $\alpha=3$
     & $-37.0^{+5.6}_{-5.7}$ & $+0.9^{+8.6}_{-8.6}$ & $+35.2^{+12.6}_{-12.7}$
     &\nodata
     & $7.29^{+1.61}_{-1.61}$ & $+0^{+1750}_{-1300}$ \\
  & $\alpha=2$
     & $-38.0^{+5.7}_{-5.6}$ & $-0.5^{+8.6}_{-8.6}$ & $+33.6^{+12.7}_{-12.6}$
     &\nodata
     & $7.25^{+1.61}_{-1.61}$ & $+100^{+1800}_{-1300}$ \\
  & $\alpha=1.7$
     & $-38.3^{+5.6}_{-5.7}$ & $-1.0^{+8.6}_{-8.6}$ & $+33.0^{+12.7}_{-12.6}$
     &\nodata
     & $7.25^{+1.61}_{-1.61}$ & $+150^{+1800}_{-1300}$ \\
\enddata
\tablenotetext{a}{This row assumes that there are no sources in
the data}
\end{deluxetable}

\begin{deluxetable}{lcc}
\tablecaption{Comptonization and Peculiar Velocity Uncertainties
for MS0451 (Nov 2000)\label{tab:sumuncert}}
 \tablewidth{0pt}
 \tablehead{
 \colhead{Uncertainty} & \colhead{$y_0 \times 10^{4}$} & \colhead{$v_p$ (km
 s$^{-1}$)}
 }
 \startdata
 Statistical: & $3.06^{+0.83}_{-0.83}$ & $-300^{+1950}_{-1250}$ \\
 & & \\
 Systematic: & & \\
 \phm{space} Common-Mode Atmospheric Removal & $^{+0.06}_{-0.06}$ & $^{+10}_{-10}$ \\
 \phm{space} Differential-Mode Atmospheric Removal & $^{+0.01}_{-0.04}$ & $^{+50}_{-25}$ \\
 \phm{space} Position Offset & $^{+0.12}_{-0.00}$ & $^{+0}_{-0}$ \\
 \phm{space} Primary Anisotropies & $^{+0.05}_{-0.05}$ & $^{+380}_{-380}$ \\
 \phm{space} Sub-millimeter Galaxies & $^{+0.20}_{-0.20}$ & $^{+650}_{-650}$
 \\
& & \\
Total:$^a$ & 3.06$^{+0.83}_{-0.83}$$^{+0.25}_{-0.21}$ & $-300^{+1950}_{-1250}$$^{+755}_{-755}$ \\
\enddata
\tablecomments{$^a$ The first number is the statistical
uncertainty, the second is the systematic uncertainty}
\end{deluxetable}

\begin{deluxetable}{rcccccl}
\tablecaption{The SuZIE sample of peculiar velocity measurements
\label{sec:t1}} \tablewidth{0pt}
  \tablehead{
    &
    & \colhead{Peculiar Velocity}
    & \multicolumn{2}{c}{Galactic Coordinates} & Distance\tablenotemark{b} \\
    \cline{4-5}
    \colhead{} & \colhead{Redshift} & \colhead{(km s$^{-1}$)\tablenotemark{a}} & \colhead{$l$ (deg.)}
    & \colhead{$b$ (deg.)} & ($h^{-1}$ Mpc)
    }
\startdata
  A1689 & 0.18 & \phn$+170^{+805}_{-600}$ & $313.39$        & \phs$61.10$ & \phn520 \\
  A2163 & 0.20 & \phn$+490^{+1310}_{-790}$ & \phn\phn$6.75$ & \phs$30.52$ & \phn570 \\
  A2261 & 0.22 & $-1400^{+1725}_{-1050}$ & \phn$55.61$       & \phs$31.86$ & \phn630 \\
  A2390 & 0.23 & $+1950^{+6275}_{-2675}$ & \phn$73.93$       & $-27.83$    & \phn650 \\
  A1835 & 0.25 & \phn$-225^{+1650}_{-1225}$ & $340.38$       & \phs$60.59$ & \phn710 \\
 ZW3146 & 0.29 & \phn$-650^{+3350}_{-1875}$ & $239.39$       & \phs$47.96$ & \phn810 \\
 MS0451 & 0.55 & \phn$+750^{+1500}_{-1125}$ & $201.50$       & $-27.32$    & 1440 \\
 0016 & 0.55 & $-4050^{+2900}_{-1775}$ & $112.55$         & $-45.54$    & 1440 \\
\enddata
 \tablenotetext{a}{Statistical uncertainties only.  An additional systematic
 uncertainty
 of $\pm750$\,km\,s$^{-1}$ is assumed for the analysis in \S\ref{sec:flow_calc}}

 \tablenotetext{b}{For a $\Omega_m=0.3$, $\Omega_\Lambda=0.7$
cosmology}
\end{deluxetable}

\begin{deluxetable}{lccccl}
\tablecaption{Limits to bulk flows in specific directions.
\label{sec:t2}} \tablewidth{0pt}
  \tablehead{
   & \multicolumn{2}{c}{Direction (Gal.)}
    &  \multicolumn{2}{c}{Velocity (km s$^{-1}$)} & \\
    \cline{2-5}
    \colhead{Flow name} & \colhead{$l$ (deg.)} & \colhead{$b$ (deg.) }
    & \colhead{Best Fit} & \colhead{95\% conf. limit} & \multicolumn{1}{c}{Ref.}}
\startdata
  CMB Dipole & 276 & +33 & $+60\pm690$\phn  & 1410 & 1 \\
  ACIF       & 343 & +52 & $+130\pm510$ \phn & 1130 & 2 \\
  LP10K      & 272 & +10 & $+40\pm930$ \phn & 1860 & 3 \\
  SMAC       & 260 & +1\phn & \phn$0\pm980$  & 1930 & 4 \\
\enddata
\tablerefs{(1) \citet{kogut94}, (2) \citet{1994ApJ...425..418L},
(3) \citet{1999ApJ...522..647W}, (4) \citet{1999ApJ...512L..79H}}
\end{deluxetable}

\clearpage

\end{document}